\documentclass[aps,prd,twocolumn,showpacs,amsmath,amssymb,nofootinbib, showpacs, showkeys]{revtex4-2}
\usepackage{graphicx}
\usepackage{color}
\usepackage{times}
\usepackage{inputenc}
\usepackage{bm}
\usepackage{ulem}
\usepackage{multirow}
\usepackage{float}
\usepackage{url}
\usepackage{mathrsfs}
\usepackage{physics}
\usepackage{comment}
\usepackage[table,xcdraw]{xcolor}
\usepackage{booktabs,makecell}
\usepackage{comment}
\usepackage[colorlinks=true,citecolor=blue,urlcolor=blue,linkcolor=blue]{hyperref}
\usepackage[caption=false]{subfig}

\usepackage{booktabs}
\usepackage{multirow}

\usepackage{graphicx}
\usepackage{dcolumn}
\usepackage{bm}

\begin{document}

\preprint{APS/123-QED}

\title{Multi-Physics Bayesian Analysis  of Neutron Star Crust Using Relativistic Mean-Field Model}
\author{Vishal Parmar$^{1}$}
\email{vishal.parmar@pi.infn.it}
\author{Ignazio Bombaci$^{1,2}$}
\affiliation{\it $^{1}$ INFN, Sezione di Pisa, Largo B. Pontecorvo 3, I-56127 Pisa, Italy}
\affiliation{\it $^{2}$ Dipartimento di Fisica, Universit\`{a} di Pisa, Largo B.  Pontecorvo, 3 I-56127 Pisa, Italy}

\date{\today}
\begin{abstract}
We study the properties of neutron-star crust within a Bayesian framework based on a unified relativistic mean-field (RMF) description of dense matter. The analysis focuses on the posterior distributions of crust properties, constrained by nuclear experimental data, chiral effective field theory, and multimessenger neutron-star observations. In the inference, the outer crust is fixed using the AME2020 nuclear mass table, supplemented by Hartree--Fock--Bogoliubov mass models, while the inner crust is described using a compressible liquid-drop model consistently coupled to the RMF interaction. The same RMF framework is used to describe the uniform core, ensuring a unified treatment across all density regimes. From the resulting posteriors, we extract  key crustal observables, including the crust--core transition density and pressure, crust thickness, crust mass, and the fractional crustal moment of inertia. We find that the transition density is primarily governed by the symmetry-energy slope $L$ and curvature $K_{\rm sym}$ evaluated at sub-saturation densities, while the transition pressure plays a central role in determining global crustal properties. The inner-crust equation of state reflects a collective interplay between isovector nuclear-matter properties rather than a dependence on any single parameter. We also assess the impact of using matched crust--core constructions and show that they can introduce systematic differences in predicted neutron-star properties when compared with fully unified treatments.
\end{abstract}

\maketitle


\section{\label{sec:level1} Introduction}

Neutron stars (NSs)  provide a natural laboratory for exploring the properties of strongly interacting dense matter under conditions far beyond the reach of terrestrial experiments. \cite{Annala2020, Annala2023, Huth2022, Tsang2024, Marino2024}. Formed in the collapse of massive stellar cores \cite{Couch_2017}, or in an electron capture supernova (ECSN) \cite{Jones_2016, PARMAR2026100470}, NSs confine matter under extreme  conditions, producing a highly stratified interior. The outer layers consist of a solid Coulomb lattice of nuclei embedded in an electron gas; deeper within, the inner crust hosts neutron-rich clusters immersed in a sea of dripped neutrons and electrons, while the core forms a uniform liquid of baryons and leptons. \cite{Haensel_2008, Haensel_2009, BKS_2015}. Although the crust contributes only about one percent of the stellar mass, it typically spans ten percent of the radius and therefore strongly influences observable quantities such as pulsar glitches \cite{Pizzochero2017}, crustal oscillation modes \cite{Kozhberov_2020}, thermal relaxation, and the cooling evolution \cite{POTEKHIN2025116, Page_2004}. A precise understanding of the crust equation of state (EOS) and of the crust--core transition is thus essential for accurately determining NS structure and assessing the impact of the crust on observable stellar properties \cite{Chamel_2016, Roca_2008, Pearson_2018}.

However, theoretical uncertainties begin already in the outer crust where  the absence of measurements for very neutron--rich isotopes introduces unavoidable model dependence \cite{Parmar_crust, BPS}. Deeper inside, the inner crust poses an even greater theoretical challenge, as nuclear clusters can assume a variety of  geometries under conditions inaccessible to terrestrial experiments, making theoretical modeling essential \cite{Avancini_2008, Avancini_2009}. In the literature, the inner crust has been predominantly described within the spherical Wigner--Seitz (WS) approximation, where the crystalline lattice is replaced by independent spherical cells \cite{Chamel2008}. While this approach has proven highly useful and remains the standard framework, it introduces geometric and boundary--condition simplifications that become increasingly inaccurate as the density approaches the crust--core transition. In particular, the WS construction cannot fully account for lattice effects, non--spherical cluster correlations, or the onset of spatial uniformization, leading to systematic uncertainties in the predicted transition density and pressure \cite{Chamel_2013, Chamel_2016, Oertel_2017}. At still higher densities, within the uniform core, the situation remains far from settled: the composition may range from nucleonic matter to the appearance of additional baryonic degrees of freedom or even transitions towards deconfined quark matter, leading to a wide spectrum of possible EOS \cite{Annala2023, Oertel_2017}. As a result, the global structure of a NS, from the surface to the center, is highly sensitive to the assumed microphysics across the entire density range, from subsaturation crustal densities to the supranuclear regime deep in the core. This sensitivity has become increasingly important in light of modern multimessenger constraints, including tidal deformability limits from GW170817 \cite{Abbott_2017, Abbott_2018}, precise mass and radius measurements of massive pulsars such as  PSR J0740+6620 \cite{Cromartie_2019} ,   PSR J0348+0432 \cite{Antoniadis_2013} and PSR J1614–2230 \cite{Demorest_2010}, PSR J0437–4715 \cite{Choudhury_2024}, and multiple theoretical/experimental data on nuclear matter (NM) properties \cite{Tsang2024} which together significantly narrow the range of viable NS  matter properties.

While the properties of the uniform core have been extensively studied, the crust remains comparatively less explored because of its greater complexity and the theoretical uncertainties associated with nonuniform matter. General--purpose EOS for supernova and proto--NS simulations often rely on nuclear statistical equilibrium at finite temperature, combined with energy--density functionals and Thomas--Fermi--type approximations in spherical WS cells \cite{Oertel_2017}.  For cold, catalyzed NSs,  several strategies have been employed to determine the structure of the NS  crust within WS approximation \cite{Chamel_2016, Carreau_2019, Carreau_2020, Oertel_2017, klausner2025neutronstarcrustinformed, Burrello_2025}. Approaches based on the thermodynamical or dynamical spinodal \cite{Ducoin_2011} provide useful insight into the onset of instabilities and can estimate the crust--core transition density, but they do not yield the full composition or geometry of the crust. Methods such as Thomas--Fermi (TF) or extended  Thomas-Fermi (ETF), that makes full
use of the nuclear-physics-informed finite-size terms of
the energy functional \cite{klausner2025neutronstarcrustinformed}  offer a detailed description but are computationally demanding when applied over wide parameter ranges. Within this landscape, compressible liquid--drop models (CLDMs), developed in \cite{Baym_1971} have emerged as an efficient and remarkably accurate framework for crust calculations. In a CLDM model, the energy of a WS  cell is expressed through bulk, surface, Coulomb, and curvature contributions, with optional shell corrections \cite{Carreau_2019, Carreau_2020, Parmar_crust}. When the surface and curvature terms are carefully calibrated to nuclear masses and semi--infinite NM, the model reproduces detailed Thomas--Fermi and density--functional results  while remaining computationally inexpensive \cite{klausner2025neutronstarcrustinformed}.  More microscopic methods, such as quantum molecular dynamics simulations \cite{PhysRevLett.102.191102}, band structure calculations \cite{CHAMEL2005109, PhysRevC.105.045807}, and quantum Monte Carlo approaches \cite{Oertel_2017}, provide valuable insight into the microscopic structure of pasta phases and into transport and superfluid properties.  
These methods, however, are extremely demanding computationally. 

Unlike the uniform core, whose EOS has been extensively explored within Bayesian and statistical frameworks, the NS  crust has received far less attention in this regard. This gap is significant because key crustal properties, such as the crust--core transition density and pressure, the crustal thickness, and the fractional moment of inertia, are highly sensitive to the microphysics of nonuniform matter and cannot be reliably quantified without a consistent treatment of uncertainties. In  the literature, the crust is constructed by stitching together different models for the outer crust, inner crust, and core, but such matching procedures inevitably introduce inconsistencies since the underlying nuclear interaction is not the same across all regions \cite{Fortin_2016, Suleiman_2021, Davis_2024, Ferreira_2020}.  Recently, efforts have focused on enforcing consistency between the crust and core through meta-modelling techniques, notably via numerical tools such as \texttt{CUTER}, which enable the matching of a nuclear-physics–informed crust to an arbitrary high-density EOS \cite{Davis2025}. A physically coherent description requires that the same nuclear energy--density functional be used from the surface to the center. Only recently have a few studies begun to address this issue by examining the statistical dependence of crustal properties on empirical NM parameters.  While the majority of such work rely on the CLDM approach \cite{Carreau_2019, Dinh_2021, Burrello_2025}, most recent work utilize the extended Thomas-Fermi approach based on a wide set of Skyrme functionals, derived from previous nuclear physics inferences and then identifies EOSs which satisfy astrophysical constraints. \cite{klausner2025neutronstarcrustinformed}. All such different calculations show variations in the inner crust EOS and derived crust-core (CC) transition points. Some of the work also attempts to study the CC transition density \cite{Ducoin_2011, Douchin_2001, Vinas2017_APhysPolB} not necessarily within a statistical framework. 

In this context, it becomes essential to construct the NS  crust EOS within a fully Bayesian and  unified relativistic framework. In this work, we perform a  Bayesian analysis of a unified RMF--based EOS that consistently treats the  inner crust, and uniform core using the same underlying nuclear interaction. We treat the parameters of the RMF Lagrangian as random variables constrained by  empirical saturation properties, nuclear experimental constraints, and multimessenger NS  observations. The outer crust is computed using the AME2020 mass table \cite{Huang_2021} supplemented with Hartree-Fock-Bogoliubov (HFB ) \cite{Samyn_2002} method using the accurately calibrated Brussels-Montreal \cite{Eya2017} energy-density functionals, such as BSk26 \cite{hfb2426}. The HFB formalism offers a precise and reliable microscopic method for computing nuclear masses, particularly for nuclei far from stability. The inner crust is obtained by combining the RMF bulk interaction with a compressible liquid--drop description of the WS  cell. For each sample of the posterior, we refit the surface parameters of the CLDM to the AME2020 atomic mass evaluations.  Motivated by the recent Bayesian study of EOS \cite{Tsang2024} that combines diverse nuclear experimental constraints with astronomical data to avoid over-reliance on any single measurement, we explore how crustal properties, such as the crust--core transition density and pressure, crustal thickness, and crustal moment of inertia, are constrained by current data. We analyze correlations between these quantities and empirical NM parameters, identify combinations most tightly constrained by observations, and provide updated estimates for the transition density and associated best--fit EOS. In addition, we quantitatively assess how individual NM properties influence the crust EOS itself across sub-saturation densities, rather than focusing solely on the transition point. By analyzing correlations and sensitivities, we identify the NM combinations most relevant for shaping the crust EOS and provide updated estimates of the transition density and associated best-fit EOS. Finally, we compare our unified RMF--based Bayesian results with recent Bayesian crust studies \cite{klausner2025neutronstarcrustinformed, Burrello_2025, Dinh_2021} in order to assess the robustness of crust--core transition predictions across different theoretical frameworks.

The paper is organized as follows. In Sec.~\ref{eos}, we outline the formalism used to construct the outer crust, inner crust, and liquid core of the NS, and briefly summarize the RMF framework and the NS observables of interest. In Sec.~\ref{eos_inference}, we describe the nuclear-physics constraints, the construction of the unified EOS, and the Bayesian inference methodology. The results are presented and discussed in Sec.~\ref{results}. Finally, Sec.~\ref{conclusion} provides a summary and outlook.

\section{\label{eos} Equation of State}

\subsection{Relativistic Mean Field Theory}

The relativistic mean-field (RMF) model provides a covariant effective-field-theory
description of dense baryonic matter in which nucleons interact through classical
mean fields associated with the scalar $\sigma$, vector $\omega_\mu$, and
isovector--vector $\vec{\rho}_\mu$ mesons \cite{Serot_1986, Chen_2014, Shen_2011, Dutra_2012, Dutra_2014, Todd_2005}. The RMF Lagrangian is given by
\begin{equation}
\begin{aligned}
\mathcal{L} =&\;
\bar{\psi}\Big[\gamma^\mu (i\partial_\mu - g_\omega \omega_\mu 
- \tfrac{1}{2} g_\rho \vec{\tau}\!\cdot\! \vec{\rho}_\mu )
- (M - g_\sigma \sigma)\Big]\psi \\
&+ \tfrac{1}{2}(\partial_\mu\sigma\partial^\mu\sigma - m_\sigma^2\sigma^2)
- \tfrac{\kappa}{3!}g_\sigma^3\sigma^3
- \tfrac{\lambda}{4!}g_\sigma^4\sigma^4 \\
&- \tfrac{1}{4}\omega_{\mu\nu}\omega^{\mu\nu}
+ \tfrac{1}{2}m_\omega^2\omega_\mu\omega^\mu
+ \tfrac{\zeta}{4!} g_\omega^4(\omega_\mu\omega^\mu)^2 \\
&- \tfrac{1}{4}\vec{\rho}_{\mu\nu}\!\cdot\!\vec{\rho}^{\,\mu\nu}
+ \tfrac{1}{2}m_\rho^2 \vec{\rho}_\mu\!\cdot\!\vec{\rho}^{\,\mu}
+ \Lambda_v g_\omega^2 g_\rho^2 (\omega_\mu\omega^\mu)
  (\vec{\rho}_\mu\!\cdot\!\vec{\rho}^{\,\mu}),
\end{aligned}
\end{equation}

where $\psi$ denotes the nucleon isodoublet, $\sigma$ is the scalar meson field responsible for medium-induced attraction, $\omega_\mu$ is the isoscalar--vector field generating short-range repulsion, and $\vec{\rho}_\mu$ is the isovector--vector field governing the response to neutron--proton asymmetry. The corresponding coupling constants ($g_\sigma$, $g_\omega$, $g_\rho$) determine, respectively, the effective nucleon mass, the saturation density and symmetric-matter pressure, and the symmetry energy. The nonlinear $\sigma$ self-interactions ($\kappa,\lambda$) regulate the incompressibility of NM \cite{Boguta_1977, MULLER_1996}, the $\omega$ self-interaction ($\zeta$) softens the high-density EOS and affects the maximum NS mass \cite{BODMER1991703}, and the isoscalar--isovector coupling $\Lambda_v$ controls the density dependence of the symmetry energy through the $L$ and $K_{\rm sym}$ parameters \cite{Horowitz_2001}. Further discussion of their separate and joint effects on the NS EOS and structure is available in the literature \cite{Fortin_2017, PhysRevC.94.035804, Fattoyev_2010, Alam_2016, Reinhard_1989}. The detailed procedure to solve the Euler-Lagrange equations and determining pressure and energy is documented in literature and can be found is \cite{Dutra_2014}.

\subsection{Outer crust}
In the outer crust, the total energy of a charge–neutral WS   cell at a given baryon density \(\rho_b\) can be written as \cite{Haensel_2008}
\begin{equation}
E_{WS}(A,Z,\rho_b)=E_N(A,Z)+E_L+E_{zp}+E_e,
\end{equation}
where \(E_N(A,Z)=M(A,Z)\) is the nuclear rest--mass contribution. The lattice and zero--point terms are given by \cite{carreau2020modeling}
\begin{equation}
E_L=-C_M\frac{(Ze)^2}{R_N}, \qquad
E_{zp}=\frac{3}{2}\hbar\omega_p\,u,
\end{equation}
with \(C_M=0.89593\) the BCC Mandelung constant, \(u=0.51138\) \cite{Chamel_2016}, and 
\(R_N=(3/4\pi n_N)^{1/3}\) with $n_N = 1/V_{WS}$. The electron contribution is \(E_e=\mathcal{E}_e V_{WS}\), where \(V_{WS}\) is the WS-cell volume.

To determine the equilibrium composition, we follow the BPS prescription \cite{BPS_1971} and identify, at fixed pressure (P), the nucleus \((A,Z)\) that minimizes the Gibbs free energy,
\begin{equation}
G(A,Z,P)=\frac{\mathcal{E}_{WS}+P}{\rho_b},
\end{equation}
where \(\mathcal{E}_{WS}=E_{WS}/V_{WS}\) and \(\rho_b=A/V_{WS}\).

\subsection{Inner crust}
With increasing density, nuclei become progressively less bound, and at the neutron--drip point, unbound neutrons begin to populate the continuum. In the inner crust, each WS cell contains a nuclear cluster embedded in an ultrarelativistic electron gas and a surrounding neutron fluid. The energy of such a configuration can be written as \cite{Pearson_2018, BPS_1971}

\begin{equation}
\label{wsenergyic}
E_{WS}=M_i(A,Z)+E_e+V_{WS}(\varepsilon_g+n_gM_n),
\end{equation}
where $M_i(A,Z)$ is the mass of the cluster written as
\begin{equation}
    \label{clustermass}
    M_i(A,Z)=(A-Z)M_n + Z M_p + E_{cl}-V_{cl}(\varepsilon_g+n_g M_n),
\end{equation}
where $M_n$, and $M_p$ are the masses of neutron and proton respectively. ${\varepsilon}_g$, and $n_g$ are the energy density and density of the neutron gas, respectively. $V_{cl}$ and $V_{WS}$ are the volumes corresponding to the cluster and the WS cell.   The energy of the cluster at a given isospin asymmetry ($\alpha$) and density ($n_0$) reads
\begin{equation}
    \label{ecluster}
    E_{cl}=E_{bulk}(n_0,\alpha)A+E_{surf}+E_{curv}+E_{coul},
\end{equation}
where $E_{surf}$, $E_{curv}$, and $E_{coul}$ are surface, curvature and Coulomb energy respectively. While the Coulomb energy within the WS approximation is well known \cite{Chamel2008}, the surface energy becomes the most important parameter within CLDM.  Considering the cluster to be spherical, the surface energy is defined as
\begin{equation}
  E_{surf} = 4\pi R_0^2A^{2/3}\sigma(\alpha),\label{eq:esurf}
\end{equation}
where $R_0 = (3/4\pi n_0)^{1/3}$ is related to the cluster density $n_0$, and
$\sigma(\alpha)$ is the nuclear surface tension that depends on the isospin asymmetry of the cluster.  We use the parametrization of surface tension proposed by Ravenhall \textit{{\it {\it et al.}}}~\cite{Ravenhall1983} which is obtained by fitting Thomas-Fermi and Hartree-Fock numerical values as,
\begin{equation}
  \sigma(\alpha) = \sigma_0\frac{2^{p+1} + b_s}{Y_p^{-p} + b_s + (1 -
  Y_p)^{-p}},\label{eq:sigma}
\end{equation}
where, $\sigma_0,p,b_s$ are the free parameters and $Y_p$ is the proton fraction inside the cluster. Similar to surface energy, the curvature energy plays an important part in describing the surface and is written as \cite{Newton_2012} 
\begin{equation}
  E_{curv} = 8\pi R_0A^{1/3}\sigma_c.\label{eq:ecurv}
\end{equation}
Here $\sigma_c$ is the curvature tension related to the surface tension
$\sigma$ as \cite{carreau2020modeling, Newton_2012},
\begin{equation}
  \sigma_c =
  \sigma\frac{\sigma_{0,c}}{\sigma_0}\gamma(\beta-Y_p),\label{eq:sigmac}
\end{equation}
with $\gamma=5.5$  and $\sigma_{0,c}, \beta$ are the parameters which  along with the $\sigma_0$ and $b_s$ needs to be fitted for a given EOS with the available experimental AME2020 mass table \cite{Huang_2021} at a fixed value of $p$. The equilibrium composition of inhomogeneous matter in the inner crust is obtained by minimizing the energy of WS cell per unit volume at a given baryon density ($\rho_b=\rho_n+\rho_p$), where $\rho_n$ and $\rho_p$ represent the neutron and proton density respectively. We use the variational method used in \cite{Carreau_2019,Newton_2012} where the Lagrange multipliers technique is used so that the auxiliary function to be minimized reads as \cite{Carreau_2019, Carreau_2020}
\begin{equation}
    \label{eq:auxillaryfunctio}
    \mathscr{F}(A,I,\rho_0,\rho_g,\rho_p)=\frac{E_{WS}}{V_{WS}}-\mu_b\rho_b,
\end{equation}
where $\mu_b$ is the baryonic chemical potential.
A detailed formulaion of the CLDM formalism  can be referred from \cite{Carreau_2019, Dinh_2021, Parmar_crust, carreau2020modeling}. In this work we use the same formalism as used \footnote{\url{https://github.com/thomascarreau/NSEOS}} in \cite{Carreau_2019, Dinh_2021} modifying it for the case of RMF bulk matter EOS.

\subsection{Liquid core}
As the density increases, the solid inner crust eventually gives way to the uniform liquid core.  
In this region, the energy density of homogeneous matter in $\beta$--equilibrium and charge neutrality \cite{NKGb_1997}, is
\begin{equation}
    \label{coreenergy}
    \varepsilon_{\mathrm{core}}
    =\varepsilon_{B}(\rho_b,\alpha)
    +\varepsilon_{e}(\rho_e)
    +\varepsilon_{\mu}(\rho_\mu),
\end{equation}
where $\varepsilon_{B}$ denotes the baryonic contribution.  The crust--core transition is identified from the crust side as the density at which the energy density of the inner--crust WS cell becomes equal to that of uniform npe\(\mu\) matter,  
\begin{equation}
\label{eq:cctransition}
    \varepsilon_{WS}(\rho_t)
    = \varepsilon_{npe\mu}(\rho_t).
\end{equation}

\vspace{0.2cm}

\section{\label{eos_inference} Bayesian Inference}
In the Bayesian approach, the posterior distribution $P(\boldsymbol{\theta}\,|\,D,H)$
quantifies the probability of the model parameters $\boldsymbol{\theta}$ after
incorporating the data $D$, and is obtained from
\begin{equation}
P(\boldsymbol{\theta}\,|\,D,H)
= \frac{L(D\,|\,\boldsymbol{\theta},H)\,P(\boldsymbol{\theta}\,|\,H)}
       {P(D\,|\,H)} .
\end{equation}
Here, $L(D\,|\,\boldsymbol{\theta},H)$ is the likelihood measuring the agreement between
the model and the data, $P(\boldsymbol{\theta}\,|\,H)$ is the prior encoding our
pre-existing knowledge of the parameters, and $P(D\,|\,H)$ is the evidence ensuring
proper normalization. We use this framework to combine nuclear and astrophysical
constraints and obtain statistically consistent posteriors for the RMF parameter
space.

\subsubsection{Priors}

In our Bayesian analysis, the model parameters $\boldsymbol{\theta}$ correspond to the
coupling constants of the RMF Lagrangian, including the meson--nucleon couplings,
nonlinear self-interaction strengths, and isoscalar--isovector mixing terms. These
parameters determine the saturation properties, symmetry energy, and high-density
behavior of the EOS. We adopt uniform  priors over broad
intervals that encompass the range of physically reasonable RMF parametrizations.
The prior ranges used in this work are summarized in Table~\ref{tab:priors}.

\begin{table}[h!]
\centering
\caption{Uniform prior ranges adopted for the RMF coupling constants in the Bayesian analysis. All quantities are dimensionless, with the exception of $k$, which has units of MeV.}
\label{tab:priors}
\begin{ruledtabular}
\begin{tabular}{lcc}
\textbf{Parameter} & \textbf{Min} & \textbf{Max} \\
\hline
$g_\sigma$      & 8.0   & 13.0  \\
$g_\omega$      & 9.0   & 14.0  \\
$g_\rho$        & 9.0   &  14.0  \\
$\kappa$        & $0.0$ & 0.05  \\
$\lambda$       & $-0.05$ & 0.05 \\
$\zeta$         & 0.0   & 0.06 \\
$\Lambda_v$     & 0.0   & 0.06 \\
\end{tabular}
\end{ruledtabular}
\end{table}

\begin{table*}[t]
\caption{\label{tab:constraints}
Summary of the empirical, theoretical, and observational constraints used in this work.}
\begin{ruledtabular}
\begin{tabular}{lcccc}
\multicolumn{5}{c}{\textbf{(i) Nuclear matter properties}} \\[2pt]
Constraint & Density $\rho$ (fm$^{-3}$) & Observable & Value & Ref. \\[3pt]

\textit{Symmetric NM (SNM)} \\
HIC (DLL)            & 0.32            & $P_{\rm SNM}$ (MeV/fm$^{3}$) & $10.1 \pm 3.0$ & \cite{Danielewicz_2002} \\
HIC (FOPI)           & 0.32            & $P_{\rm SNM}$ (MeV/fm$^{3}$) & $10.3 \pm 2.8$ & \cite{LEFEVRE2016112} \\
GMR                   & 0.16            & $K_{\rm sat}$ (MeV)          & $230 \pm 30$   & \cite{Rutel_2005}  \\[4pt]

\textit{Symmetry energy and symmetry pressure} \\
$\alpha_D$            & 0.05            & $J(\rho)$ (MeV)                 & $15.9 \pm 1.0$ & \cite{alpha_d} \\

\textit{Nuclear masses} \\
Mass (Skyrme)         & $0.101\pm0.005$ & $J(\rho)$ (MeV)                 & $24.7 \pm 0.8$ & \cite{Brown_2013} \\
Mass (DFT)            & $0.115\pm0.002$ & $J(\rho)$ (MeV)                 & $25.4 \pm 1.1$ & \cite{PhysRevC.85.024304} \\
IAS                   & $0.106\pm0.006$ & $J(\rho)$ (MeV)                 & $25.5 \pm 1.1$ & \cite{DANIELEWICZ2017147} \\[4pt]

\textit{Heavy-ion collisions (isospin-sensitive)} \\
HIC (Iso-diff)        & $0.035\pm0.011$ & $J(\rho)$ (MeV)                 & $10.3 \pm 1.0$ & \cite{Tsang_2009} \\
HIC (n/p ratio)       & $0.069\pm0.008$ & $J(\rho)$ (MeV)                 & $16.8 \pm 1.2$ & \cite{MORFOUACE2019135045} \\
HIC ($\pi$ ratio)     & $0.232\pm0.032$ & $J(\rho)$ (MeV)                 & $52 \pm 13$    & \cite{PhysRevLett.126.162701} \\
HIC (n/p flow)        & $0.240$         & $P_{\rm sym}$ (MeV/fm$^3$)   & $12.1 \pm 8.4$ & \cite{RUSSOTTO2011471} \\[10pt]

\multicolumn{5}{c}{\textbf{(ii) $\chi$EFT constraints at low density}} \\[2pt]
$\chi$EFT PNM band & $\rho = 0.04$--$0.20$ fm$^{-3}$ & $E_{\rm SNM}(n)$, $E_{\rm PNM}(n)$ & 10\% expansion of  Ref.~\cite{Drischler_2016} & \cite{Drischler_2016} \\[10pt]

\multicolumn{5}{c}{\textbf{(iii) Astrophysical constraints}} \\[2pt]
Constraint & $M (M_\odot)$ & $R$ (km) & $\Lambda$ & Ref. \\[3pt]

\textit{Gravitational-wave observations} \\
GW170817 (LIGO/Virgo) & 1.4 &  & $190^{+390}_{-120}$ & \cite{Abbott2017} \\[4pt]

\textit{NICER pulsars} \\
PSR J0030+0451 & $1.34^{+0.15}_{-0.16}$ & $12.71^{+1.14}_{-1.19}$ & & \cite{Riley_2019} \\
 PSR J0030+0451 & $1.44^{+0.15}_{-0.14}$ & $13.02^{+1.24}_{-1.06}$ & & \cite{Miller_2019} \\
 PSR J0740+6620 & $2.07^{+0.07}_{-0.07}$ & $12.39^{+1.30}_{-0.98}$ & & \cite{Riley_2021} \\
 PSR J0740+6620 & $2.08^{+0.07}_{-0.07}$ & $13.70^{+2.6}_{-1.5}$    & & \cite{Miller2021} \\
PSR J0437--4715 & $1.418^{+0.037}_{-0.037}$ & $11.36^{+0.95}_{-0.63} $  & & \cite{Choudhury_2024} \\

\end{tabular}
\end{ruledtabular}
\end{table*}

\subsubsection{Constraints}

Motivated by the recent analysis of the NS EOS by Tsang \textit{et al.}~\cite{Tsang2024}, we adopt a similarly diverse set of constraints that combines nuclear-experimental inputs, empirical saturation properties, NICER radius measurements, and gravitational-wave tidal deformabilities. Using such complementary observables avoids reliance on any single dataset and provides balanced coverage from sub-saturation to supranuclear densities. In our RMF-based Bayesian framework, we further include the low-density chiral EFT band and an additional pulsar mass--radius constraint. The full set of constraints employed in this work is summarized below.

\textit{\textbf{(i)-Nuclear matter properties}:-}
 In the symmetric NM (SNM) sector, we take the standard Giant Monopole Resonance (GMR) motivated value
\(K_{0} = 230 \pm 30~\mathrm{MeV}\) at \(\rho_0 \simeq 0.16~\mathrm{fm^{-3}}\) \cite{Rutel_2005}. At supra-saturation densities, the pressure of SNM is constrained at $\approx 2\rho_0$ by flow analyses of relativistic Au+Au collisions, denoted HIC(DLL) \cite{Danielewicz_2002} and HIC(FOPI) \cite{LEFEVRE2016112}, which provide consistent values for $P_{\mathrm{SNM}}(2\rho_0)$ with uncertainties at the level of a few MeV/fm$^3$. The density dependence of the symmetry energy $J(\rho)$ is constrained using a set of
complementary observables that probe well-defined effective densities. At low
densities around $\rho \simeq 0.05~\mathrm{fm^{-3}}$, the electric dipole
polarizability ($\alpha_D$)  of  ${}^{208}\mathrm{Pb}$ constrains the symmetry energy  \cite{alpha_d}. In the region
$\rho \simeq 0.10$--$0.12~\mathrm{fm^{-3}}$, independent information from
nuclear-mass systematics (Skyrme and DFT analyses) and isobaric analogue 
states further restricts the symmetry energy \cite{Brown_2013, PhysRevC.85.024304, DANIELEWICZ2017147}. Sub-saturation constraints from 
heavy-ion reactions, including isospin diffusion and neutron/proton spectral 
ratios, extend coverage up to $\rho \sim 0.4\,\rho_0$ \cite{Tsang_2009, MORFOUACE2019135045}. At higher densities approaching 
and slightly exceeding $n_0$, charged-pion spectral ratios and neutron/proton 
elliptic-flow differences provide direct constraints on the symmetry pressure \cite{PhysRevLett.126.162701, RUSSOTTO2011471}. 
Taken together, these measurements form a coherent and density-resolved set of 
constraints on both $J(\rho)$ and $P_{\mathrm{sym}}(n)$ across the range relevant 
for the NS crust and outer core.

\textbf{(ii)- $\chi$EFT constraints at low density}
We use constraints from chiral effective field theory calculations for symmetric NM and pure neutron matter by Drischler \textit{et al.} \cite{Drischler_2016}, obtained using many body perturbation theory with several chiral Hamiltonians.  To account for theoretical uncertainties and to remain compatible with other \textit{ab initio} calculations, we conservatively expand the $\chi$EFT uncertainty bands by 10 percent. We do not apply pressure constraints, since the density derivative of the energy per nucleon is subject to significant uncertainties at these densities, as discussed in \cite{Carreau_2019}.

\textbf{(iii)- Astrophysical constraints:}

\textbf{Gravitational-wave observations:}
We incorporate constraints from the binary NS merger GW170817, which provides sensitivity to the high-density behavior of the EOS through tidal deformability measurements. For each EOS candidate, stellar masses, radii, and tidal deformabilities are computed self-consistently by solving the Tolman–Oppenheimer–Volkoff equations together with the linear tidal-perturbation equations \cite{2008ApJ...677.1216H}, allowing direct comparison with the GW170817 inferred tidal response. We do not include GW190425 in our analysis, since its large mass asymmetry and weaker tidal information do not provide competitive constraints on the EOS.

\textbf{Mass–radius measurements:} We further constrain the EOS using mass-radius (M-R) information from three well-studied pulsars, PSR J0030+0451 \cite{Riley_2019, Miller_2019}, PSR J0740+6620 \cite{Riley_2021, Miller2021}, and PSR J0437--4715 \cite{Choudhury_2024}. For PSR J0030+0451 and PSR J0740+6620, NICER pulse-profile modeling provides joint constraints on stellar mass and radius, with PSR J0740+6620 probing the higher-mass regime through its precisely measured radio-timing mass of $M = 2.08 \pm 0.07~M_\odot$ \cite{2021ApJ...915L..12F}. PSR J0437--4715 complements these measurements by sampling the lower-mass region, which is more sensitive to the EOS near saturation density. For all three sources, we construct likelihoods in the M-R plane using kernel density estimation of the published posterior samples and incorporate them directly into our Bayesian analysis. 

The complete set of empirical, experimental, and observational inputs employed in our Bayesian analysis is summarized in Table~\ref{tab:constraints}. We additionally constrain the nuclear saturation density to $\rho_0 = 0.153 \pm 0.005~\mathrm{fm^{-3}}$ and the  energy per nucleon at saturation to $e_0 = -16.1 \pm 0.2~\mathrm{MeV}$, and require that every EOS in our ensemble supports a maximum NS mass exceeding $2\,M_\odot$.

\begin{table*}
\caption{\label{tab:posteriors_eos}
Posterior median values and credible intervals for nuclear--matter (NM) parameters at saturation density, namely the incompressibility $K_0$, symmetry energy $J$, slope of the symmetry energy $L$, and curvature of the symmetry energy $K_{\rm sym}$, together with neutron--star (NS) observables including the maximum TOV mass $M_{\max}$, the radius and tidal deformability of a canonical $1.4\,M_\odot$ neutron star ($R_{1.4}$ and $\Lambda_{1.4}$), the dimensionless moment of inertia $I_{1.4}$, and the amoment of inertia relevant for PSR~J0737--3039A $I_{1.338}$ . Each entry reports the posterior median along with the 68\% and 95\% credible intervals (CI).}
\begin{ruledtabular}
\begin{tabular}{lccc}
{Quantity} & {68\% CI} & {Median} & {95\% CI} \\[3pt]
\hline
\multicolumn{4}{c}{{Nuclear-matter parameters}} \\[4pt]

$K_0$ (MeV)
  & 231.28 -- 266.07 
  & 251.82 
  & 208.54 -- 276.19 \\

$J$ (MeV)
  & 31.996 -- 35.139 
  & 33.361 
  & 30.801 -- 37.170 \\

$L$ (MeV)
  & 61.118 -- 76.106 
  & 67.041 
  & 56.753 -- 89.122 \\

$K_{\rm sym}$ (MeV)
  & $-66.15$ -- $-14.99$ 
  & $-43.44$ 
  & $-79.78$ -- 17.94 \\[8pt]

\hline
\multicolumn{4}{c}{{NS observables }} \\[4pt]

$M_{\rm max}$ ($M_\odot$)
  & 2.030 -- 2.233 
  & 2.102 
  & 2.004 -- 2.422 \\

$R_{1.4}$ (km)
  & 12.760 -- 13.123 
  & 12.933 
  & 12.610 -- 13.347 \\

$\Lambda_{1.4}$
  & 551.57 -- 667.75
  & 603.85
  & 511.63 -- 744.37 \\

$I_{1.4}/MR^2$
  & 0.3497 -- 0.3570
  & 0.3539
  & 0.3448 -- 0.3591 \\

$I_{1.338}\,[10^{45}\,\mathrm{g\,cm^2}]$
  & 1.509 -- 1.586
  & 1.545
  & 1.479 -- 1.632 \\

\end{tabular}
\end{ruledtabular}
\end{table*}

\section{\label{results} RESULTS AND DISCUSSIONS}

To probe the posterior distribution with high reliability, we rely on the nested-sampling Monte Carlo framework MLFriends \cite{Buchner_2016, Buchner_2019}, accessed through the UltraNest package \cite{Buchner2021}. UltraNest employs the slice-sampling  \cite{2019MNRAS.483.2044H}, which maintains stable convergence even as the dimensionality of the problem increases. The total number of sampling steps is set adaptively: repeated nested-sampling iterations are performed until the Bayesian evidence, $\log Z$, stabilizes to within the required tolerance. This strategy ensures that both posterior estimation and evidence evaluation remain robust throughout the analysis. In total, we generate $\sim 40\,000$ unified EOSs,
each of which spans the full density range from the outer crust to several times nuclear
saturation density. All constraints listed in Table~\ref{tab:constraints} are applied
directly to these unified EOSs, ensuring that the inference consistently incorporates
both the low-density nuclear physics and the high-density astrophysical information.

\subsection{Posterior distribution of EOS and NS properties}
We begin by discussing our results for the NM properties and the EOS, focusing on their posterior distributions obtained from the Bayesian analysis. The resulting posterior distributions for the nuclear-matter (NM) parameters at
saturation density, as well as key NS observables, are summarized in
Table~\ref{tab:posteriors_eos}.

\begin{figure}
    \centering
    \includegraphics[width=1\linewidth]{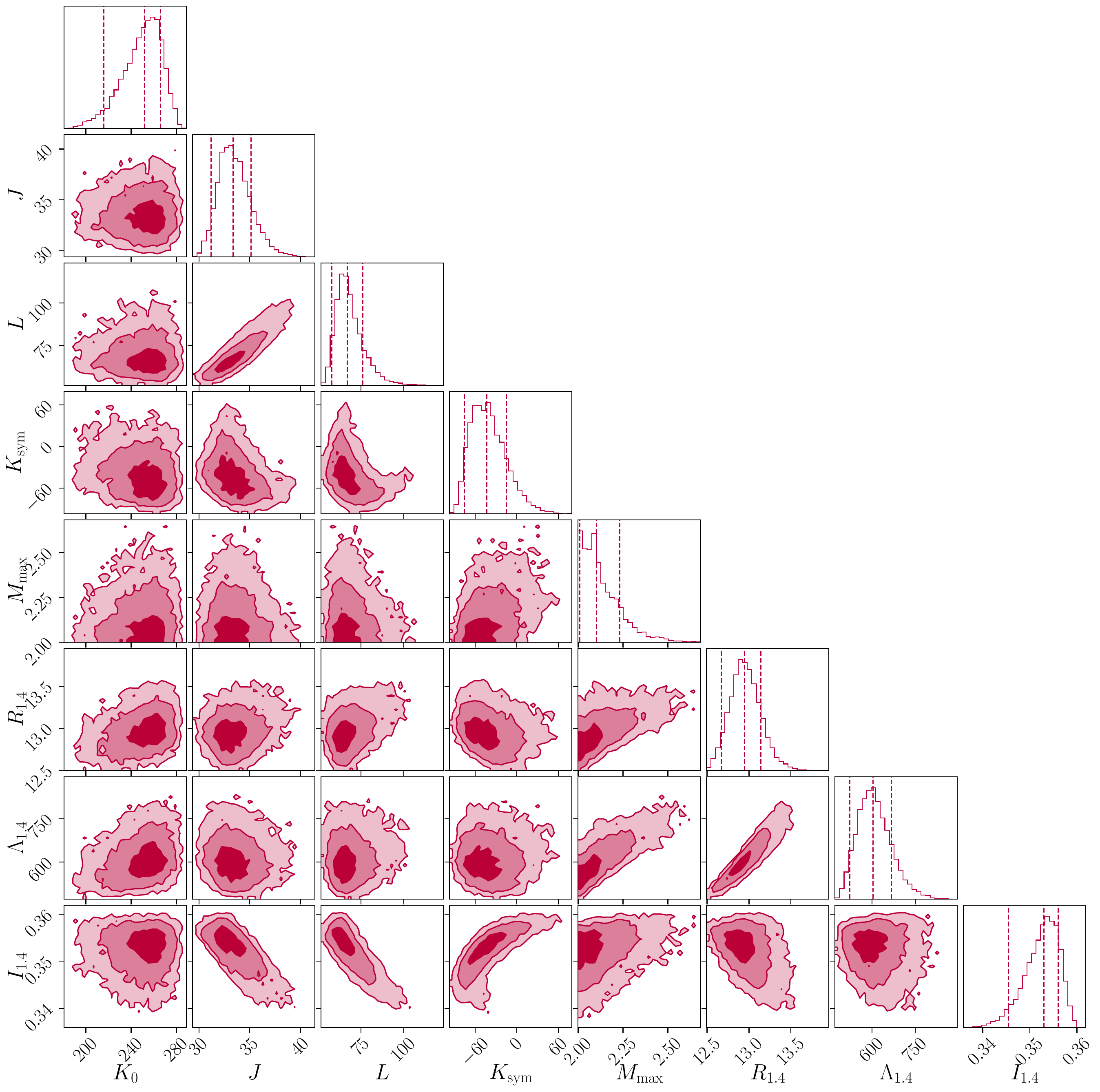}
    \caption{
Marginalized posterior distributions of the nuclear--matter parameters 
$K_0$, $J$, $L$, and $K_{\mathrm{sym}}$, and the NS  properties 
$M_{\max}$, $R_{1.4}$, $\Lambda_{1.4}$, and $I_{1.4}$. 
The diagonal panels show the one--dimensional marginalized posteriors, with 
vertical lines marking the 68\% confidence intervals (CIs). 
The off--diagonal panels display the two--dimensional joint posteriors, where 
the filled contour ellipses correspond to the 1$\sigma$, 2$\sigma$, and 
3$\sigma$ CIs. 
Darker shades indicate regions of higher probability, while lighter shades 
represent the broader, lower--probability extensions of the posterior distribution.
}

    \label{fig:eos_posterior}
\end{figure}

Our inferred NM properties are compatible with several existing studies that combine gravitational-wave and NICER constraints on the NS EOS. In particular, comparison with the recent analysis of Tsang \textit{et al.}~\cite{Tsang2024} shows overall consistency between the posterior ranges, despite differences in the underlying modeling strategy. While that work adopts a meta-model framework and reports slightly higher preferred values of the symmetry energy parameters $J$ and $L$ with $34.9^{+1.7}_{-2.0}$ and $83.6^{+19.2}_{-16.8}$ MeV, our results remain consistent within uncertainties. We further find that the symmetry incompressibility $K_{\mathrm{sym}}$ in our RMF-based analysis tends to favor more negative values, as also found in \cite{adil_2025}, whereas the meta-model posteriors of Ref.~\cite{Tsang2024} are approximately symmetric between positive and negative regions. Our inferred incompressibility $K_{\mathrm{sat}}$ is slightly higher compared to the values reported by Tsang \textit{et al.}~\cite{Tsang2024}, which is typical for RMF-based models once NICER and gravitational-wave constraints, together with the $2 M_\odot$ maximum-mass requirement, are imposed \cite{Zhu_2023, adil_2025}, while remaining well within the allowed range.

For NS properties, we find that the maximum mass spans a relatively wide range, extending up to about $2.5,M_\odot$, reflecting residual uncertainties in the high-density EOS. The predicted radius of a $1.4,M_\odot$ star is in excellent agreement with the analysis of Tsang \textit{et al.} \cite{Tsang2024}  which finds the radius to be in the range $12.4$–$13.3$ km. In our case, the corresponding credible intervals are $12.60$–$13.24$ km at 95\% credibility and $12.76$–$13.12$ km at 68\% credibility. The tidal deformability $\Lambda_{1.4}$ is also consistent with both the results of Tsang \textit{et al.}   \cite{Tsang2024} and the constraints inferred from GW170817. In addition, we report the normalized moment of inertia, for which we obtain a 95\% credible interval of $0.3448 < {I}_{1.4} < 0.3591$, with a median value of ${I}_{1.4} = 0.3539$. 
Furthermore, PSR~J0737--3039A \cite{Kramer_2006} is the primary component of the double--pulsar system, with a precisely measured mass of $M=1.338\,M_\odot$. Owing to relativistic spin--orbit coupling in the binary, it is expected that a direct measurement of its moment of inertia will become feasible in the near future, providing a unique and independent probe of the NS EOS \cite{Miao_2022}. In Table~\ref{tab:posteriors_eos}, we report the inferred moment of inertia of PSR~J0737--3039A, characterized by a median value and a corresponding 95\% credible interval. When compared with existing theoretical predictions, such as those summarized in Table~1 of \cite{Miao_2022}, our median value is found to be slightly higher than several earlier hadronic and hybrid--EOS estimates obtained from non-unified constructions. Nevertheless, it is in good agreement with the result of Silvia  \textit{et al.} \cite{Silva_2021} and remains well within the overall range spanned by contemporary EOS models. A future precise measurement of the moment of inertia of PSR~J0737--3039A would therefore provide a powerful and independent constraint on the NS EOS, with sensitivity not only to the core properties but also to the structure of the crust through its contribution to the stellar radius and mass distribution.

The full joint posterior distribution, including correlations between the EOS parameters and NS properties, is shown in the corner plot in Fig.~\ref{fig:eos_posterior}. A strong correlation between the symmetry energy $J$ and its slope parameter $L$ is clearly visible, reflecting their coupled role in determining the isovector sector of the EOS. Among NS observables, the radius $R_{1.4}$ and the tidal deformability $\Lambda_{1.4}$ show a pronounced correlation, as expected from their common dependence on the pressure at intermediate densities. In contrast, the incompressibility $K_0$ exhibits no strong correlation with the NS properties considered here. Of all stellar observables, the moment of inertia shows the strongest sensitivity to isovector parameters, being substantially affected by the symmetry energy, its slope $L$, and the curvature parameter $K_{\mathrm{sym}}$. While $J$ and $L$ display an approximately linear negative correlation, $K_{\mathrm{sym}}$ tends to show a mild positive correlation with the moment of inertia.

\begin{figure}
    \centering
    \includegraphics[width=1\linewidth]{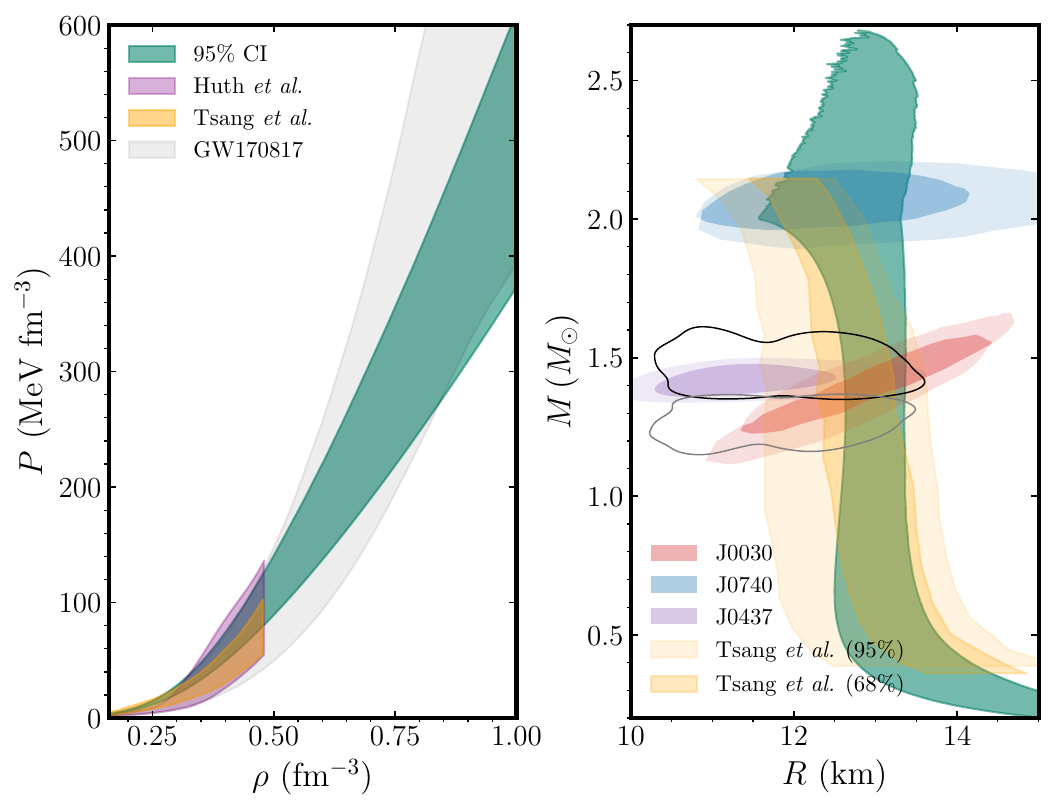}
    \caption{
Left panel: Pressure as a function of baryon density for the ensemble of EOS consistent with all applied constraints. The shaded green band shows the 95\% credible interval obtained from our Bayesian analysis. For comparison, we also display the constraints of Huth \textit{et al.}~ \cite{Huth2022} and the multi-physics constrained band reported by Tsang \textit{et al.}~ \cite{Tsang2024}, together with the pressure constraints inferred from GW170817~\cite{Abbott2017}. 
Right panel: Mass--radius relations corresponding to the same EOS ensemble. The green band denotes the 95\% credible region obtained from the TOV solutions. Shaded elliptical regions show the mass--radius constraints from NICER pulse-profile modeling for PSR J0030+0451 and PSR J0740+6620, with darker and lighter regions indicating the 68\% and 95\% credibility intervals, respectively, while PSR J0437--4715 is shown as an additional low-mass constraint. The contours correspond to the GW170817 mass--radius posterior. 
}
    \label{fig:eos_mr}
\end{figure}

Figure~\ref{fig:eos_mr} summarizes the EOS and M-R constraints obtained from our Bayesian analysis of unified EOS within RMF framework. The inferred pressure–density band is compatible with gravitational-wave constraints from GW170817 and with representative nuclear-theory results, including those of Huth \textit{et al.} \cite{Huth2022} and Tsang \textit{et al.} \cite{Tsang2024}, over the full density range shown. In the M-R plane, our EOS ensemble lies largely within the 95\% and 68\% credible regions reported by Tsang \textit{et al.}, indicating broad consistency with their analysis. For the most massive NSs, our models predict slightly larger radii than the median Tsang bands, while remaining compatible within uncertainties.

\subsection{Crust Properties}
Having established an ensemble of EOSs that is consistent with a broad set of nuclear, gravitational-wave, and X-ray observations, we now turn to the central focus of this work. In the following, we use this vetted EOS ensemble to investigate the properties of the NS crust and its interface with the core. 
\begin{figure}
    \centering
    \includegraphics[width=1\linewidth]{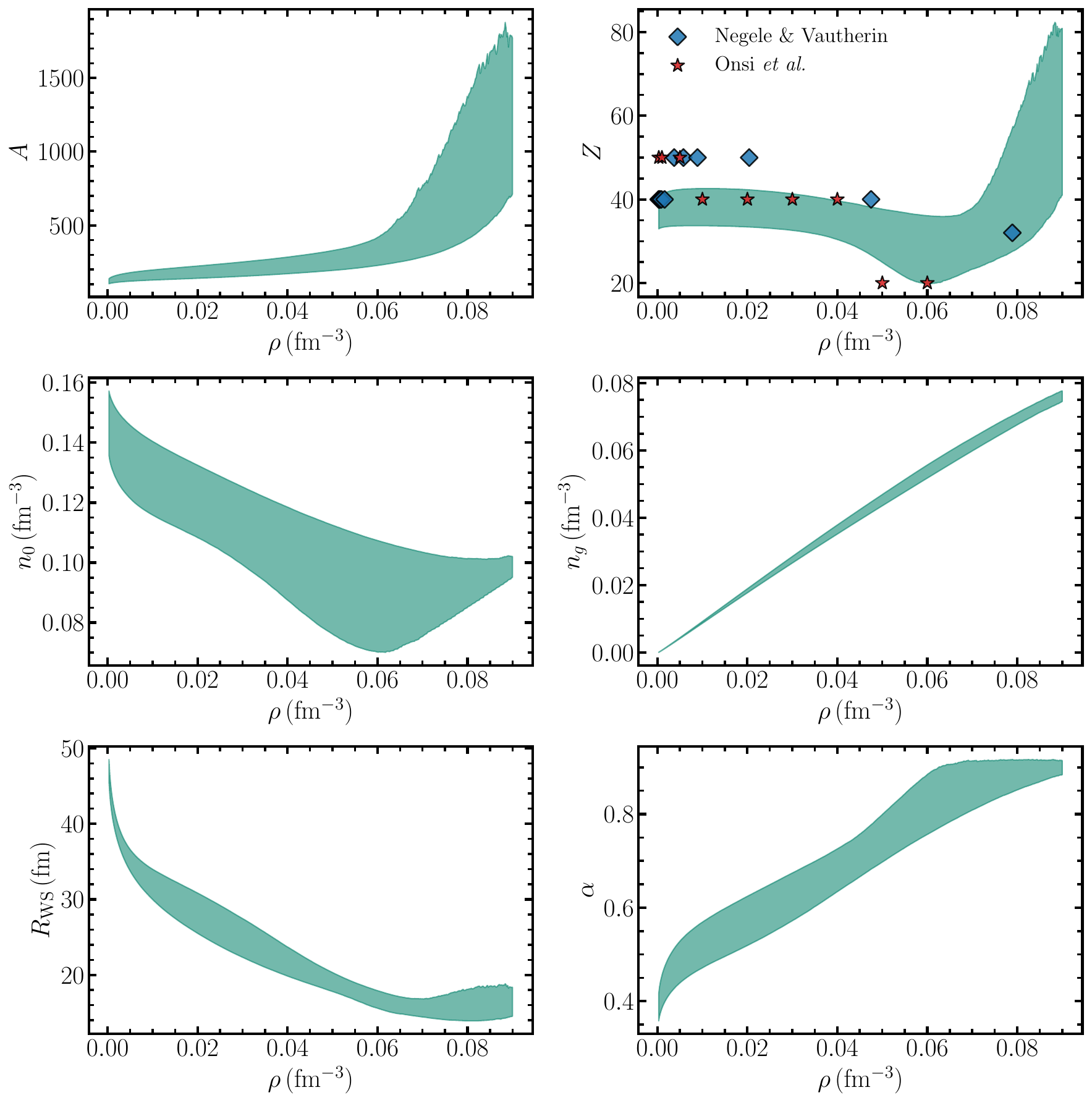}

    \caption{
Properties of nuclear clusters in the inner crust as a function of baryon density $\rho$ inside the crust.
Shown are the cluster mass number $A$ (upper left),
atomic number $Z$ (upper right),
 density of the cluster $n_0$ (middle left),
neutron gas density $n_g$ (middle right),
the WS  cell radius $R_{\mathrm{WS}}$ (lower left),
and the cluster isospin asymmetry $\alpha$ (lower right).
The quantum calculations of Negele and Vautherin~\cite{NEGELE1973298} and Onsi \textit{et al.}~\cite{Onsi_2008} are also shown.
}
    \label{fig:crust_structure}
\end{figure}
In Fig.~\ref{fig:crust_structure}, we show the main properties of nuclear clusters in the inner crust within the CLDM formalism as functions of the baryon density. Shown are the cluster mass number $A$, atomic number $Z$, density of the cluster $n_0$, the surrounding neutron gas density $n_g$, the WS  cell radius $R_{\rm WS}$, and the cluster isospin asymmetry $\alpha$. The quantum calculations of Negele and Vautherin~\cite{NEGELE1973298} and Onsi et al.~\cite{Onsi_2008} are shown for comparison for the $Z$ of the cluster. These quantities are sufficient to characterize the structure of the inner crust.

We find that $A$, $Z$, and $n_0$ are strongly model dependent, in line with earlier CLDM studies \cite{Carreau_2019, Dinh_2021} as well as ETF and \cite{klausner2025neutronstarcrustinformed}  microscopic ETFSI calculations \cite{Pearson_2018}. In the low-density region of the inner crust, $A$ increases slowly with density, while it rises more rapidly close to the crust--core transition, indicating the gradual conversion toward homogeneous matter in the core. Owing to the semi-classical nature of the CLDM, $Z$ varies smoothly with density, in contrast to fully quantum approaches such as Negele and Vautherin \cite{NEGELE1973298} or Onsi \textit{et al.} \cite{Onsi_2008}, and to ETFSI calculations where proton shell corrections produce pronounced step-like features. Nevertheless, the persistent tendency toward $Z \simeq 40$ (which corresponds to a filled proton subshell) throughout the inner crust remains consistent with these quantum results. At the same time, uncertainties arising from NM properties, experimental constraints, and astrophysical data can be larger than typical shell-correction effects, as also discussed in Ref.~\cite{Carreau_2019}. Moreover, the evolution of $A$, $Z$, and the WS radius $R_{\rm WS}$ directly affects the elastic properties of the inner crust, in particular the shear modulus. Since the shear modulus controls the frequencies of crustal shear modes, these quantities become especially relevant for the interpretation of quasi periodic oscillations observed in magnetar flares.

Among the displayed quantities, the neutron gas density $n_g$ is the least model dependent and increases monotonically with density. We also observe correlated trends, namely a continuous increase of $A$, a decrease of the cluster central density $n_0$, and a reduction of the WS  radius $R_{\rm WS}$ with increasing density, together with a steady growth of the asymmetry $\alpha$. These correlations represent quasi-universal features that have been reported in several crust studies \cite{Carson_2019, klausner2025neutronstarcrustinformed, Ding_2019, Dinh_2021} and discussed in \cite{klausner2025neutronstarcrustinformed}. As density increases, clusters become increasingly neutron rich, leading to a systematic rise of $\alpha$ across the inner crust. The behaviour of $n_0$, $A$, and $Z$ can be directly linked to the density dependence of the symmetry energy and its higher-order parameters in the sub-saturation density region, where crust physics is most relevant~\cite{Parmar_crust, Dutra_2021}. Since the inner crust consists of clusters embedded in a neutron gas, its equilibrium configuration follows from energy minimization, making the symmetry energy a key ingredient. An accurate determination of the NM symmetry energy is therefore essential for reliable crust predictions, and conversely, crust properties provide valuable constraints on the symmetry energy. We return to this point in more detail in the following sections.

\begin{figure}
    \centering
    \includegraphics[width=1\linewidth]{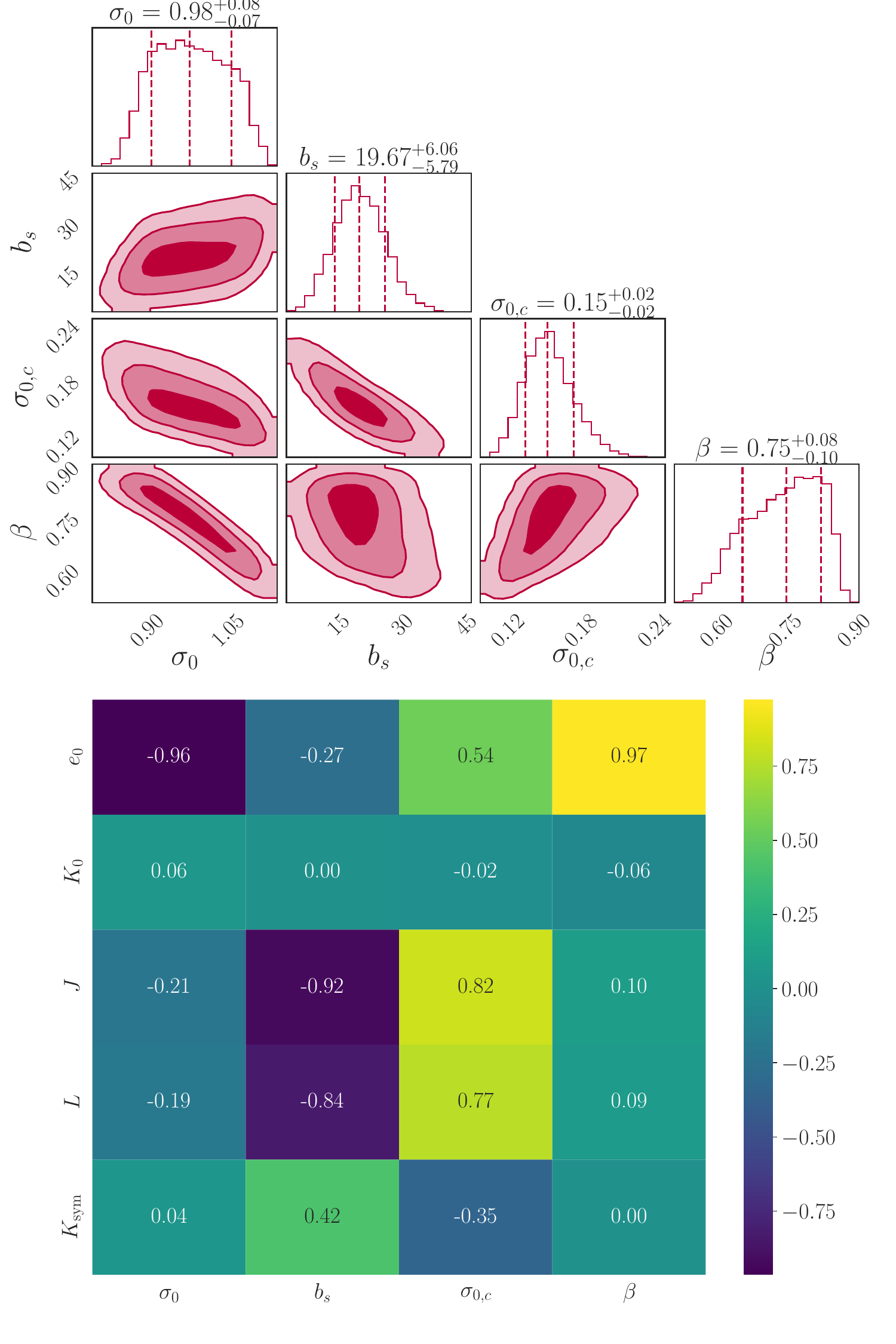}
    \caption{Posterior distributions of the surface–friction parameters and their correlations with NM properties.}

    \label{fig:surface_params}
\end{figure}

The structure of the NS inner crust is strongly controlled by the parametrization of the surface and curvature energies of nuclear clusters (see Eqs. \eqref{eq:esurf} and \eqref{eq:ecurv}). Since the surface properties of extremely neutron-rich nuclei are not accessible experimentally, we adopt a semiempirical description of the cluster energy, Eq.~(\ref{ecluster}), and calibrate the associated parameters using experimental masses from the AME2020 table \cite{Huang_2021}. The surface--curvature parameter set $\boldsymbol{S}=\{\sigma_0, b_s, \sigma_{0,c}, \beta, \gamma, p\}$ is constrained by minimizing the penalty function $\chi^2(\boldsymbol{S})$ \cite{Dobaczewski_2014}, with a systematic theoretical uncertainty of $0.1~\mathrm{MeV}$ \cite{carreau2020modeling}. The isospin-dependence parameter is fixed to $p=3$, as commonly adopted in surface-energy studies \cite{Lattimer_1991, Avancini_2009}, while the curvature parameter is set to $\gamma=5.5$ following Ref.~\cite{Newton_2012}. The resulting distributions of the optimized surface parameters for the RMF models considered in this work are  are shown in Fig.~\ref{fig:surface_params}. The dependence of the surface parameters on the NM properties is shown in the lower panel of Fig.~\ref{fig:surface_params}.

The surface tension of symmetric NM is tightly constrained to $\sigma_0 = 0.98^{+0.08}_{-0.07}$ MeV fm$^{-2}$, reflecting the strong sensitivity of nuclear binding energies to the overall magnitude of the surface term, which is therefore efficiently fixed by experimental mass data \cite{Pomorski_2003,Carreau_2019}. In contrast, the isovector surface parameter $b_s = 19.7^{+6.1}_{-5.8}$ exhibits a substantially broader posterior distribution, as nuclear masses probe only moderately neutron-rich systems and provide limited constraints on the isospin dependence of the surface energy \cite{Newton_2012,Carreau_2019}. The effective surface parameter relevant for clusterized matter near the crust–core transition, $\sigma_{0,c} = 0.15 \pm 0.02$ MeV fm$^{-2}$, is significantly reduced with respect to $\sigma_0$, indicating a pronounced softening of the surface tension in extremely neutron-rich environments, consistent with earlier CLDM and ETF analyses \cite{Newton_2012,Carreau_2019}. The asymmetry exponent $\beta = 0.75^{+0.08}_{-0.10}$ is compatible with values commonly adopted in NS crust calculations based on phenomenological surface-energy parametrizations \cite{Lattimer_1991,Avancini_2009}. Furthermore, a strong negative correlation is observed between $\sigma_0$ and $\beta$, reflecting a compensation mechanism in mass fits whereby an increase in the surface tension of symmetric matter must be accompanied by a steeper isospin dependence in order to reproduce experimental binding energies. A similar correlation is found between the isovector surface parameter $b_s$ and the curvature-related surface term.

In this analysis, the surface parameters exhibit clear correlations with bulk NM properties: the symmetric surface tension $\sigma_0$ and the asymmetry exponent $\beta$ show strong negative and positive correlations, respectively, with the saturation energy $e_0$, while the isovector surface parameter $b_s$ and the curvature-related term $\sigma_{0,c}$ display pronounced correlations with the symmetry energy $J$ and its slope $L$. This behavior contrasts with the meta-modeling study of Ref.~\cite{carreau2020modeling}, where most surface parameters were found to be only weakly correlated with NM properties, with the notable exception of a mild correlation between $e_0$ and $\sigma_0$. This further signifies that, within the CLDM framework, the use of universal or fixed surface and curvature parameters is generally not justified, as these quantities depend sensitively on the underlying nuclear-matter properties of the adopted interaction. Employing surface parameters calibrated from a different model can therefore lead to a systematic under- or over-estimation of the crust EOS and, in particular, of the crust–core transition density and pressure.

\begin{figure}
    \centering
    \includegraphics[width=1\linewidth]{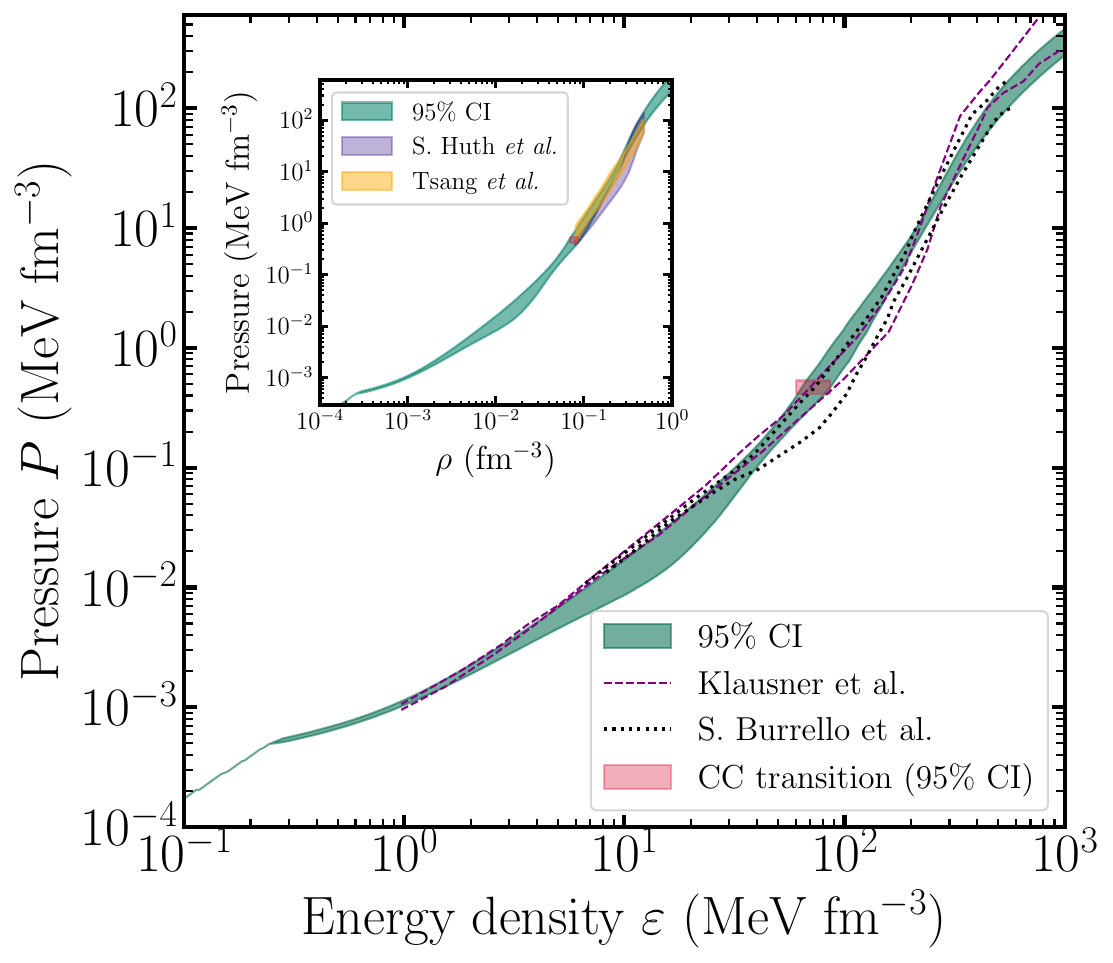}
    \caption{
Pressure as a function of energy density for the unified crust--core EOS  obtained from our Bayesian analysis, shown on logarithmic axes to span the full density range from the outer crust to the core. The shaded region denotes the $95\%$ credible interval of the posterior distribution. For comparison, we include the upper and lower bounds from Klausner \textit{et al.} \cite{klausner2025neutronstarcrustinformed} and the microscopic crust calculations of Burrello \textit{et al.} \cite{Burrello_2025}, both converted to consistent units. The inset shows the pressure as a function of baryon density in the crust region, where our results are compared with the  constraints of Huth \textit{et al.} \cite{Huth2022} and the  band of Tsang \textit{et al.} \cite{Tsang2024}. The magenta band show the CC transition uncertainty. 
}

    \label{fig:crust_eos}
\end{figure}

In Fig.~\ref{fig:crust_eos}, we present the fully unified equation of state, with particular emphasis on the neutron-star crust. In our calculations, the outer crust is fixed using the AME2020 atomic mass evaluation together with the HFB24 mass table, thereby minimizing uncertainties associated with nuclear masses. The inner crust predictions are compared with recent  crust calculations that incorporate uncertainties from NM properties and astrophysical constraints within a meta--model framework~\cite{Burrello_2025}, as well as with Skyrme-based results~\cite{klausner2025neutronstarcrustinformed}. We note that in the latter case the unified  EOS was not included in the inference procedure. The inset shows the pressure as a function of baryon density in the crust, where our posterior EOS is compared with the constraint bands reported by Huth \textit{et al.}~\cite{Huth2022} and Tsang \textit{et al.}~\cite{Tsang2024}. Our results are in excellent agreement with Tsang \textit{et al.}   \cite{Tsang2024} at low densities and show a mild deviation from the band of Huth \textit{et al.} \cite{Huth2022}. This difference can be traced back to the distinct modeling strategies: the analysis of Tsang \textit{et al.}   \cite{Tsang2024} departs from that of Huth \textit{et al.} due to its strict reliance on chiral effective field theory ($\chi$EFT) constraints below $1.5\,n_0$, which leads to a systematically softer EOS. In the present work, while we also employ $\chi$EFT constraints at low densities, we relax the corresponding band by $10\%$, resulting in an EOS that closely follows the behavior reported by Tsang \textit{et al.}~\cite{Tsang2024}. Using this ensemble of EOSs, we further compare our crust predictions with other models in the main panel. We find that the crustal EOS remains strongly model dependent across all three approaches, including the present work. While the different models show good agreement at the lowest densities of the inner crust, significant deviations emerge at intermediate densities. The overall uncertainty is largely driven by the behavior of the density dependence of the symmetry energy and its higher-order derivatives, as previously demonstrated in Ref.~\cite{Parmar_crust} and many other works on the NS crust. Within the RMF framework adopted here, which offers less flexibility in nuclear-matter properties than meta--model and Skyrme approaches, the resulting EOS is softer at intermediate inner-crust densities and exhibits a broader spread across models. Close to the crust--core transition, however, it becomes marginally stiffer. This behavior implies reduced elastic rigidity and lower shear-mode frequencies in the intermediate crust, while the increased stiffness near the transition favors a higher transition pressure and a thicker, more massive crust, with direct consequences for crustal oscillations and glitch-related observables.

\begin{figure}
    \centering
    \includegraphics[width=1\linewidth]{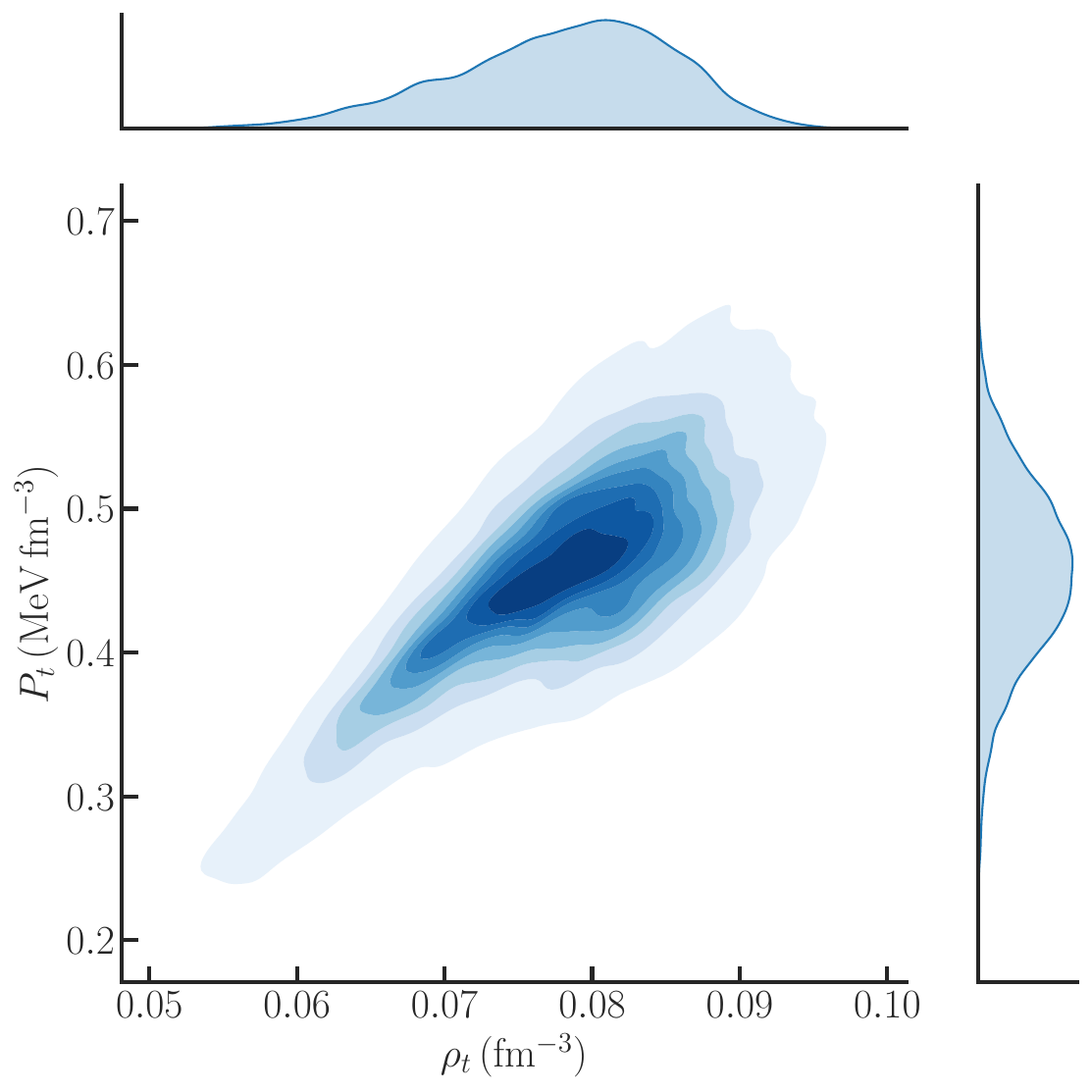}
    \caption{
Kernel density estimate (KDE) of the crust--core transition density 
$\rho_t$ and transition pressure $P_t$ obtained from the posterior 
distribution of unified NS EOS. 
The shaded contours represent the highest-posterior-density (HPD) regions, 
with darker tones indicating areas of higher probability. 
The joint distribution highlights the correlated uncertainty between 
$\rho_t$ and $P_t$, reflecting the combined influence of nuclear-matter 
properties and astrophysical constraints on the NS inner crust.
}
    \label{fig:cc_transition}
\end{figure}

Among all crustal properties, the crust--core transition point is the most critical quantity, as it governs the extent, mass, and elastic response of the neutron-star crust. Figure~\ref{fig:cc_transition} shows the posterior distributions of the crust--core transition density, $\rho_{\mathrm{t}}$, and pressure, $P_{\mathrm{t}}$, while Table~\ref{tab:cc_transition} summarizes the corresponding $95\%$ credible intervals from the present work together with results from other studies of the crust--core transition. In the present  RMF analysis, we do not find values of $\rho_{t} \lesssim 0.05~\mathrm{fm}^{-3}$ or $\rho_{t} \gtrsim 0.10~\mathrm{fm}^{-3}$. Nevertheless, the posterior distribution exhibits a low--density tail extending toward $\rho_{t} \lesssim 0.065~\mathrm{fm}^{-3}$. As shown in Table~\ref{tab:cc_transition}, our inferred values of $\rho_{\mathrm{t}}$ are comparable to those obtained using the CLDM and meta--model approaches  tuned to reproduce the  ($\chi$EFT) \cite{Burrello_2025} but estimate higher values of $P_{\mathrm{t}}$. Our $\rho_{\mathrm{t}}$ is also slightly lower than the values reported from ETF and CLDM calculations in Ref.~\cite{klausner2025neutronstarcrustinformed}, while our corresponding $P_{\mathrm{t}}$ remains broadly consistent with their results.  Taken together, Fig.~\ref{fig:cc_transition} and Table~\ref{tab:cc_transition} highlight the significant model dependence and intrinsic complexity involved in modeling the NS  crust. In the present work, where the EOS is constrained by a broad and complementary set of nuclear and astrophysical observables, our results indicate that values of $\rho_{\mathrm{t}}$ approaching or exceeding $0.10~\mathrm{fm}^{-3}$ remain incompatible with current constraints.

\begin{table}
\caption{Crust--core transition density $n_{cc}$ and pressure $P_{cc}$ 
predicted by different microscopic and phenomenological models. 
Uncertainties denote $68\%$ credible intervals.}
\label{tab:cc_transition}

\small
\setlength{\tabcolsep}{3pt}
\centering
\begin{tabular}{@{}lcc@{}}
\hline\hline
Model 
& $\rho_{t}$ [fm$^{-3}$] 
& $P_{t}$ [MeV fm$^{-3}$] \\
\hline

ETF (Skyrme)~\cite{klausner2025neutronstarcrustinformed} 
& $0.090^{+0.010}_{-0.008}$ 
& $0.510^{+0.077}_{-0.082}$ \\

CLDM (Skyrme)~\cite{klausner2025neutronstarcrustinformed} 
& $0.092^{+0.009}_{-0.007}$ 
& $0.520^{+0.055}_{-0.070}$ \\

CLDM (Meta)~\cite{Burrello_2025}
& $0.074^{+0.014}_{-0.014}$ 
& $0.277^{+0.137}_{-0.137}$ \\[.6em]

{CLDM (RMF, this work)} 
& ${0.0788^{+0.0066}_{-0.0086}}$ 
& ${0.457^{+0.059}_{-0.061}}$ \\
\hline\hline
\end{tabular}
\end{table}

\begin{figure*}
    \centering
    \includegraphics[width=.7\linewidth]{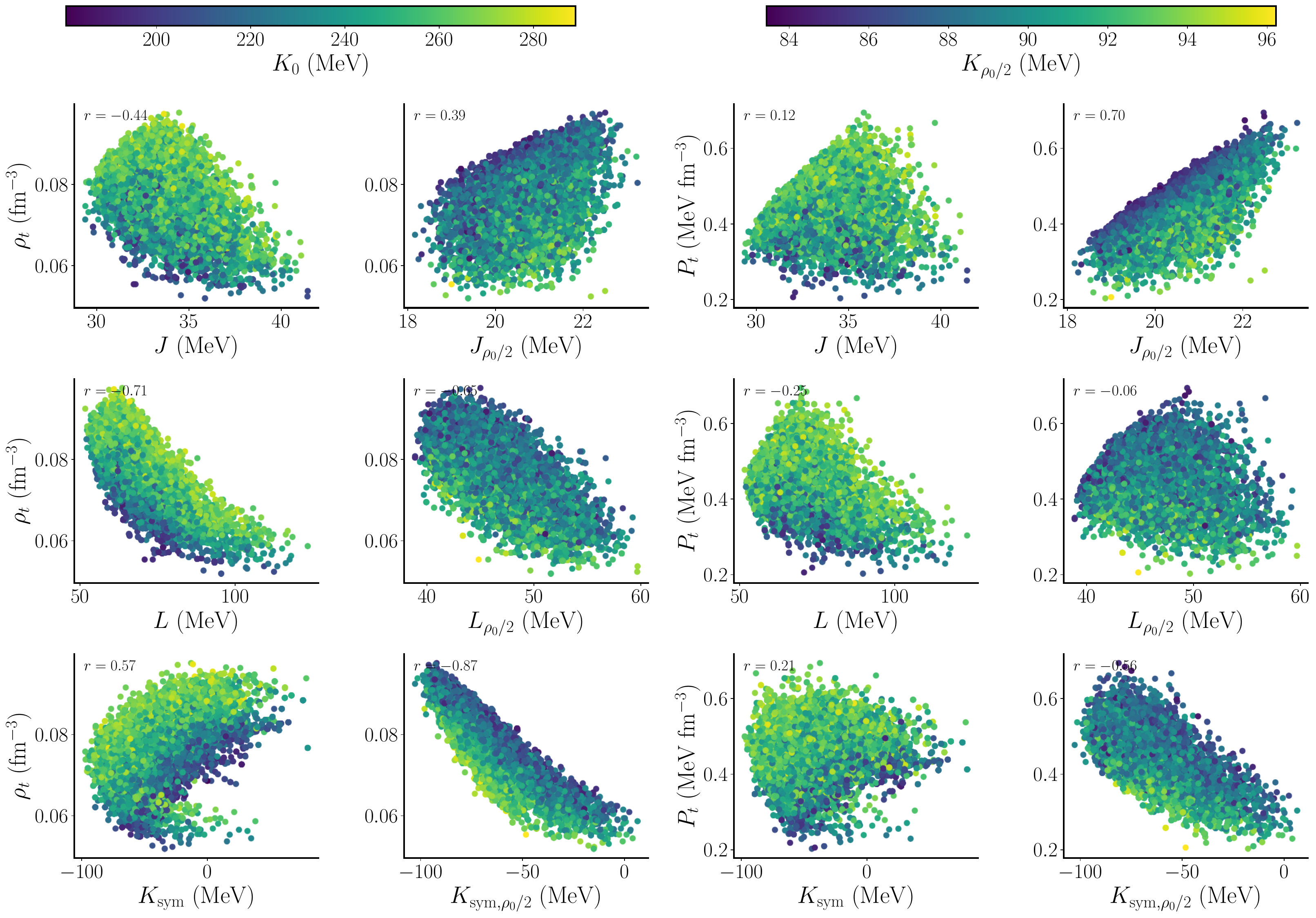}
    
    \caption{
Correlations of the crust--core transition density $\rho_t$ (left block) and transition pressure $P_t$ (right block) with NM properties evaluated at saturation density and at $\rho_0/2$. 
From top to bottom, the rows show correlations with the symmetry energy $J$, its slope $L$, and the curvature parameter $K_{\mathrm{sym}}$. 
In each row, the left (right) panel corresponds to the quantity evaluated at saturation density ($\rho_0/2$). 
The color scale indicates the incompressibility $K_0$ (left block) and $K_{\rho_0/2}$ (right block). 
Pearson correlation coefficients are reported in each panel.
}

    \label{fig:cc_correlation}
\end{figure*}

\begin{figure*}
    \centering
    \includegraphics[width=0.45\linewidth]{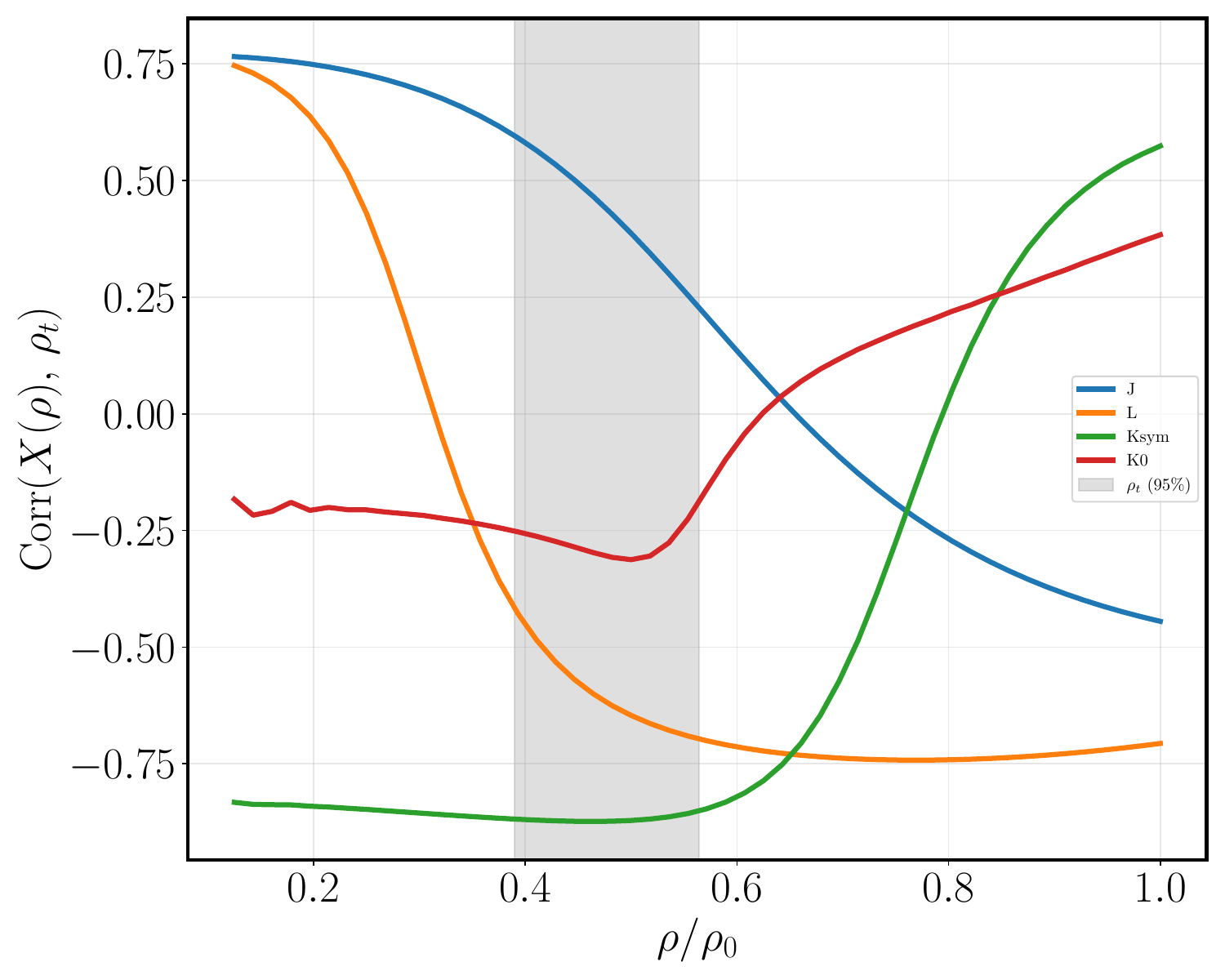}
    \includegraphics[width=0.45\linewidth]{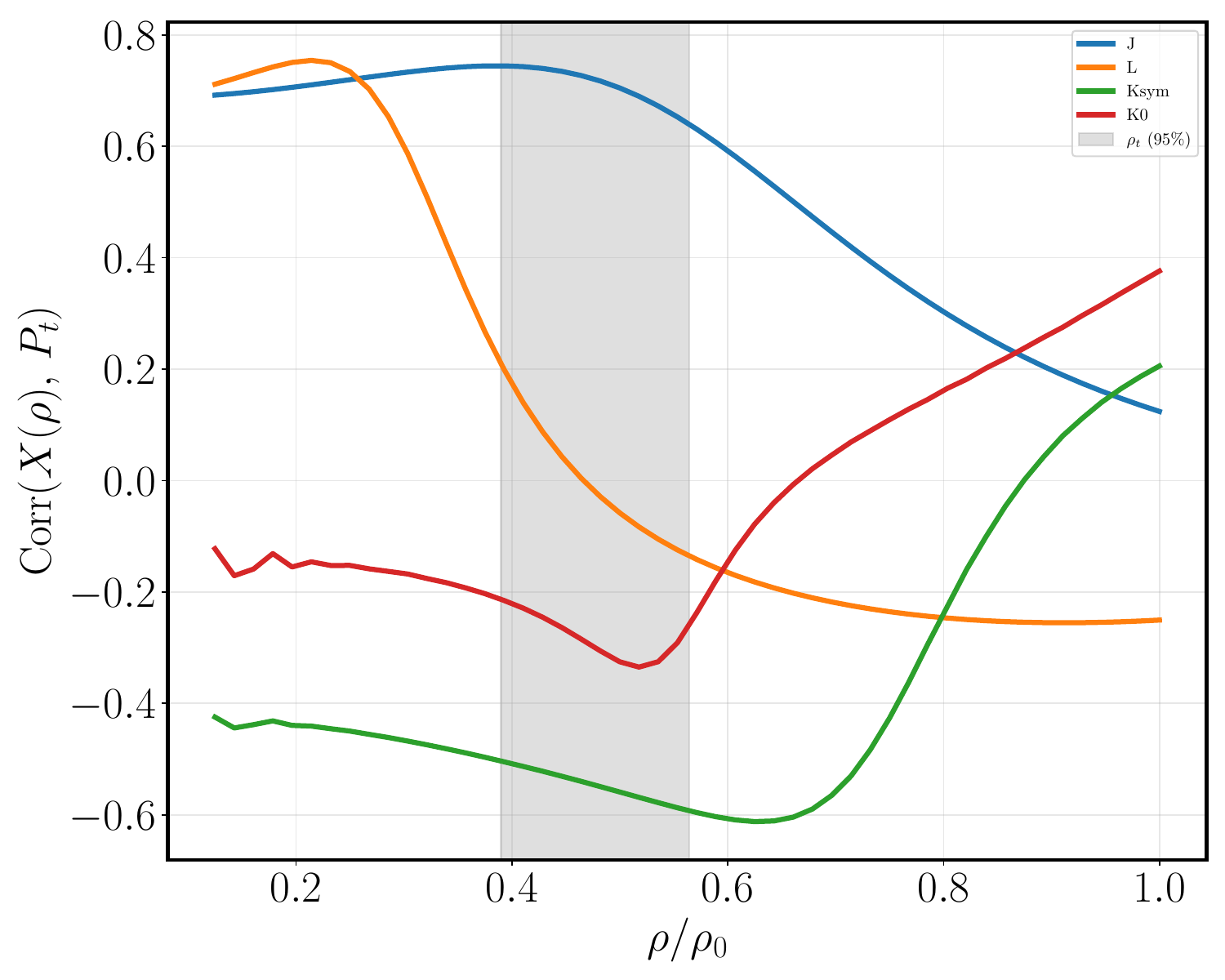}
    
    \caption{
Correlations of the crust--core transition density $\rho_t$ (left block) and transition pressure $P_t$ (right block) with NM  properties evaluated at different reference densities. 
The gray band represent the uncertainty in the CC transition density. 
}
    \label{fig:cc_tran_2}
\end{figure*}

Since CC transition point is extremely sensitive to the implied EOS, in Fig.~\ref{fig:cc_correlation}, we investigate the correlations between NM  properties and the crust--core transition density $\rho_t$ and pressure $P_t$. 
Such correlations have been extensively studied in the literature using a variety of approximations, modeling strategies, and statistical frameworks for the EOS 
\cite{Pearson_2018, Dutra_2021, Li_2020, carreau2020modeling, Carreau_2019, Parmar_crust, Balliet_2021}.
These studies have argued that, rather than the symmetry energy $J$ itself, the slope $L$ and curvature $K_{\mathrm{sym}}$ of the symmetry energy play a more important role in determining the crust--core transition
\cite{Parmar_crust, Li_2020}, motivating the present analysis. Since NS crusts reside at sub-saturation densities, it is also more appropriate to examine correlations between $\rho_t$ and $P_t$ and NM properties evaluated below saturation density, as emphasized in Ref.~\cite{Balliet_2021}.
This is further justified by the fact that most terrestrial symmetry-energy observables probe sub-saturation densities around $\rho \simeq (2/3)\rho_0$.
Accordingly, Fig.~\ref{fig:cc_correlation} is organized such that the left (right) block shows correlations with NM properties evaluated at saturation density $\rho_0$ ($\rho_0/2$), while the left (right) panels correspond to $\rho_t$ ($P_t$).

From the full Bayesian inference constrained by a broad set of nuclear and astrophysical observations, we find no significant correlation between the symmetry energy $J$ and either $\rho_t$ or $P_t$ at saturation density.
The correlation of $J$ with $P_t$ shows a strong enhancement when evaluated at $\rho_0/2$, while its impact on $\rho_t$ remains weak.
In contrast, the slope parameter $L$ exhibits a clear correlation with $\rho_t$, which remains quantitatively similar when evaluated at $\rho_{0/2}$ while changing the relationship from non-linear to linear.
The influence of $L$ on $P_t$ is generally weak in both cases. For the curvature parameter $K_{\mathrm{sym}}$, the correlation with $\rho_t$ is non-linear and relatively weak at saturation density, but becomes significantly stronger when evaluated at $\rho_0/2$, highlighting the importance of sub-saturation physics for the crust--core transition.
Overall, we find that $\rho_t$ shows stronger correlations with NM properties than $P_t$.
The color coding represents the incompressibility $K_0$, for which no significant correlation with either $\rho_t$ or $P_t$ is observed, in agreement with previous studies. This behavior highlights the complex and indirect role of incompressibility in determining crust--core transition properties. Furthermore, within the CLDM framework, when the surface and curvature energy parameters are fitted to experimental nuclear masses, we find no significant correlation between these parameters and the crust--core transition density $\rho_t$ or transition pressure $P_t$, confirming the findings of Ref.~\cite{Carreau_2019}.

In Fig.~\ref{fig:cc_tran_2}, we examine in detail the density dependence of the correlations between the crust--core transition density $\rho_t$ and transition pressure $P_t$ and nuclear-matter properties. To this end, we compute the relevant NM quantities at each density and evaluate their correlations with $\rho_t$ and $P_t$.  It is evident that the correlations of $\rho_t$ and $P_t$ are not strongest at saturation density and exhibit a complex density dependence. The symmetry energy $J$ shows a significant correlation with $\rho_t$ when evaluated at sub-saturation densities, $\rho/\rho_0 \lesssim 0.4$ (or $\rho \approx 0.06\,\mathrm{fm}^{-3}$), but this correlation rapidly weakens at higher densities. In contrast, compared to correlations evaluated at  $\rho_0$, the correlations of $\rho_t$ with the slope parameter $L$ and the curvature parameter $K_{\mathrm{sym}}$ increase substantially at sub-saturation densities, particularly in the range $\rho/\rho_0 \simeq 0.5$--$0.7$.
This density interval is especially relevant, as discussed by Tsang \textit{et al.} \cite{Tsang2024}, since NM properties, most notably $L$, are best constrained around these densities by terrestrial experiments. The incompressibility shows no meaningful correlation with $\rho_t$ or $P_t$ at any density, consistent with previous findings.
These results suggest that a combination of the slope $L$ and curvature $K_{\mathrm{sym}}$ evaluated at sub-saturation densities around $\rho/\rho_0 \simeq 0.5 - 0.7$ provides the most informative connection to the crust--core transition.
For reference, the gray band in the figure indicates the posterior range of $\rho_t$, illustrating that the strongest correlations occur near $\rho/\rho_0 \approx 0.5$--$0.7$, where NM properties are simultaneously best constrained. Compared to $\rho_t$, the transition pressure $P_t$ exhibits a stronger correlation with the symmetry energy $J$ than with the slope parameter $L$ or the curvature parameter $K_{\rm sym}$. Accordingly, while $\rho_t$ is primarily influenced by $L$ and $K_{\rm sym}$ in the density range $0.5$--$0.7\,\rho/\rho_0$, the behavior of $P_t$ in the same density region is dominated by $J$, with a secondary contribution from $K_{\rm sym}$.

\begin{table*}
\centering
\caption{Parametrizations of the crust--core transition density $\rho_t$ and their diagnostic performance.}
\begin{tabular}{l l l c c}
\toprule
toprule
Model & Input & Equation & Statistic & Value \\
\midrule

\multirow{4}{*}{ (This work)}
& \multirow{4}{*}{$L_{\rho_0/2},\,K_{\mathrm{sym},\rho_0/2}$}
& \multirow{4}{*}{$
\rho_t =
0.0654
- 1.16\times10^{-4} L_{\rho_0/2}
+ 8.55\times10^{-8} L_{\rho_0/2} K_{\mathrm{sym},\rho_0/2}^{2}
$}
& RMSE & $3.46\times10^{-3}$ \\
& & & MAE  & $2.72\times10^{-3}$ \\
& & & Mean & $3.57\%$ \\
& & & Max  & $25.14\%$ \\

\midrule
\multirow{4}{*}{Ducoin \textit{et al.} \cite{Ducoin_2011}}
& \multirow{4}{*}{$L_{\rho_0}$}
& \multirow{4}{*}{$
\rho_t =
0.0963
- 3.75\times10^{-4} L_{\rho_0}
$}
& RMSE & $9.26\times10^{-3}$ \\
& & & MAE  & $7.91\times10^{-3}$ \\
& & & Mean & $9.74\%$ \\
& & & Max  & $25.99\%$ \\

\midrule
\multirow{4}{*}{Steiner \textit{et al.} \cite{Steiner_2015}}
& \multirow{4}{*}{$L_{\rho_0}$}
& \multirow{4}{*}{$
\rho_t =
\frac{L}{{30}}\!\left(
0.1327
- 0.0898 \frac{L}{{70}}
+ 0.0228 \frac{L}{{70}}^{2}
\right)
$}
& RMSE & $6.12\times10^{-3}$ \\
& & & MAE  & $5.05\times10^{-3}$ \\
& & & Mean & $6.30\%$ \\
& & & Max  & $25.64\%$ \\

\bottomrule
\bottomrule
\end{tabular}
\label{tab:eq_crust}
\end{table*}

In the literature, several attempts have been made to derive simple algebraic relations for estimating the CC transition density, $\rho_t$, using bulk NM properties of a given EOS, most commonly evaluated at saturation density. Early work by Ducoin \textit{et al.}~\cite{Ducoin_2011} proposed a linear anticorrelation between $\rho_t$ and the symmetry energy slope parameter $L$. Subsequently, Steiner \textit{et al.}~\cite{Steiner_2015} suggested a nonlinear parametrization written in terms of rescaled symmetry energy slopes, motivated by the observed curvature in the correlation. These relations established the central role of isovector properties in controlling the transition, although their quantitative accuracy remains limited. Ducoin \textit{et al.} further showed that evaluating NM properties at subsaturation density improves the predictive power, an observation later supported by independent studies such as Ref.~\cite{Balliet_2021}.

In the present work, we systematically revisit these empirical relations using a large ensemble of RMF EOS that are simultaneously constrained by nuclear experiments and astrophysical observations.   Since the correlation between the CC transition density $\rho_t$ and symmetry energy parameters is seen to be strongest at subsaturation densities, in particular around $\rho_0/2$, it is natural to express $\rho_t$ directly in terms of the symmetry energy slope $L$ and curvature $K_{\rm sym}$ evaluated at this density. Instead of assuming a predefined polynomial structure, we searched for an empirical relation that is both physically transparent and statistically reliable. For this purpose, we employed symbolic regression using the \texttt{PySR} \footnote{\url{https://github.com/MilesCranmer/PySR}} framework, which applies an evolutionary algorithm to identify analytic expressions by balancing algebraic simplicity against predictive accuracy. A key advantage of this method is that no functional form is imposed a priori, allowing the algorithm to autonomously explore and rank candidate relations.

The input features were $(L, K_{\rm sym})$ evaluated at $\rho_0/2$, while the training targets were the RMF calculated transition densities $\rho_t$. The operator set was restricted to elementary algebraic functions to preserve physical interpretability. Across multiple evolutionary populations, several candidate expressions emerged with low complexity and strong statistical performance. Notably, relatively simple expressions were found to perform nearly as well as more elaborate ones, indicating that the dependence of $\rho_t$ on $(L, K_{\rm sym})$ is compact. The final symbolic regression model adopted in this work, along with a  few other analytic functions used in literature, is reported in Table~\ref{tab:eq_crust}, together with its diagnostic metrics. The performance comparison presented in Table~\ref{tab:eq_crust} shows that the linear relation based solely on $L$ provides the weakest description of the transition density. Allowing for nonlinear dependence in $L$ already improves the agreement, while the explicit inclusion of the symmetry energy curvature $K_{\rm sym}$ leads to a further and significant reduction in the global error measures. At the same time, the maximum percentage error remains comparable across all parametrizations. This indicates that extreme deviations arise from a small subset of EOS at the edges of parameter space and reflects the intrinsic complexity of the CC transition problem.

\begin{figure*}
    \centering
    \includegraphics[width=1\linewidth]{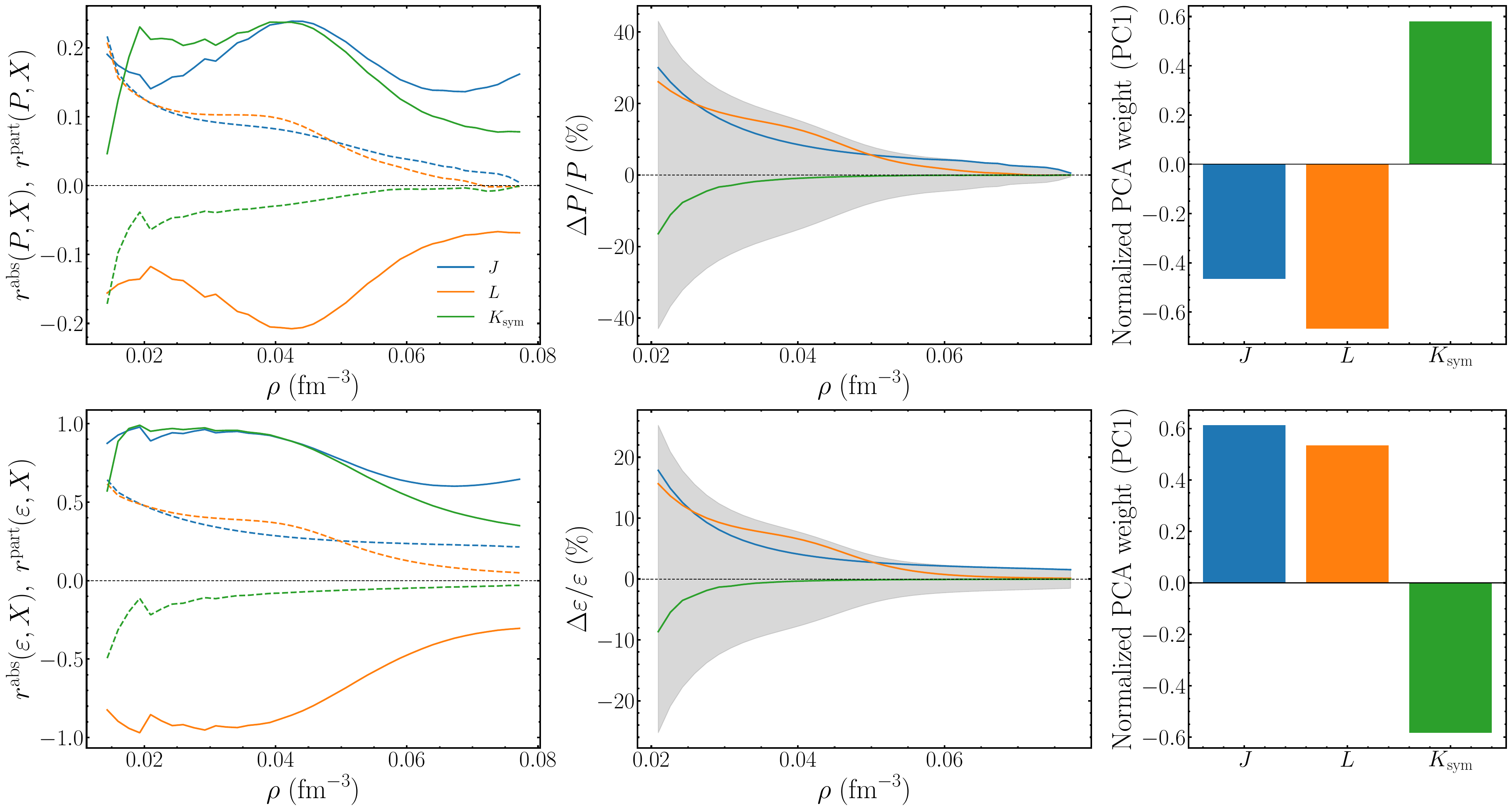}

\caption{
Sensitivity of the NS inner crust EOS to isovector
NM properties as a function of baryon density.
The left panels show the absolute (dashed lines) and partial (solid lines)
correlations between the pressure $P$ (top) and energy density $\varepsilon$
(bottom) and the symmetry-energy parameters $J$, $L$, and $K_{\rm sym}$,
all evaluated at the same density.
The middle panels display the corresponding induced fractional variations
$\Delta P/P$ (top) and $\Delta\varepsilon/\varepsilon$ (bottom).
The shaded gray bands indicate the total combined fractional uncertainty,
obtained by adding in quadrature the individual contributions from
$J$, $L$, and $K_{\rm sym}$.
The right panels show the dominant symmetry-energy mode obtained from a
principal-component analysis of the local EOS response, highlighting the
correlated combinations of $J$, $L$, and $K_{\rm sym}$ that primarily govern
the pressure and energy density in the inner crust.
}

    \label{fig:crust_sensitivity}
\end{figure*}

It has been emphasized extensively in the literature that the EOS of the NS inner crust is strongly influenced by the density
dependence of the nuclear symmetry energy. In particular, while the crust--core
transition density is seen to correlate more strongly with the symmetry-energy
slope and curvature parameters, $L$ and $K_{\rm sym}$, than with the symmetry
energy at saturation $J$, it remains important to quantify how isovector nuclear
matter properties impact the EOS throughout the inner crust. Most previous studies have examined this dependence by correlating the inner
crust EOS with NM properties evaluated at saturation density.
However, as demonstrated above, the relevant correlations emerge at subsaturation
densities characteristic of the crust. Motivated by this, we quantify the impact
of isovector NM properties on the inner crust EOS by comparing the
energy density and pressure with the symmetry-energy parameters evaluated at the
same baryon density. This procedure defines a local parameter space
$\{\rho_i, \varepsilon_i, P_i, J_i, L_i, K_{{\rm sym},i}\}$, where the nuclear
matter properties are consistently calculated at the density $\rho_i$ associated
with each EOS point. At each density, we first compute the absolute (Pearson)
correlation between the EOS quantities $P$ and $\varepsilon$ and the corresponding
isovector NM parameters. To isolate the direct sensitivity of the EOS to individual parameters and to
account for the strong mutual correlations among $J$, $L$, and $K_{\rm sym}$, we
also evaluate the partial correlation coefficients. For example, the partial
correlation between the pressure and the symmetry energy $J$, conditioned on $L$
and $K_{\rm sym}$, is defined as
\begin{equation}
r^{\rm part}(P,J) =
\mathrm{corr}\!\left(P - \hat{P}(L,K_{\rm sym}),
                     J - \hat{J}(L,K_{\rm sym})\right),
\label{eq:partial_corr}
\end{equation}
where $\hat{P}$ and $\hat{J}$ denote the linear regressions on the remaining
variables. Analogous expressions are used for the energy density. Finally, the quantitative impact of isovector NM uncertainties on the
inner crust EOS is assessed by estimating the induced fractional variations in
pressure and energy density, $\Delta P / P$ and $\Delta \varepsilon / \varepsilon$.
At each density $\rho_i$, these variations are obtained by combining the local
EOS response to independent variations of the symmetry-energy parameters
$J$, $L$, and $K_{\rm sym}$, weighted by their respective dispersions within the
posterior.

Specifically, the total fractional variation in the pressure is defined as
\begin{equation}
\left(\frac{\Delta P}{P}\right)_i
=
\frac{1}{P_i}
\left[
\begin{aligned}
&\left(\frac{\partial P}{\partial J}\sigma_J\right)^2
+
\left(\frac{\partial P}{\partial L}\sigma_L\right)^2 \\
&\qquad
+
\left(\frac{\partial P}{\partial K_{\rm sym}}\sigma_{K_{\rm sym}}\right)^2
\end{aligned}
\right]^{1/2},
\label{eq:deltaP}
\end{equation}

with an analogous expression used to compute
$\Delta \varepsilon / \varepsilon$ for the energy density.
Here, $\sigma_J$, $\sigma_L$, and $\sigma_{K_{\rm sym}}$ denote the standard
deviations of the corresponding NM parameters.

In Fig.~\ref{fig:crust_sensitivity}, we summarize the sensitivity of the NS
inner crust EOS to isovector NM properties.
The left column shows the absolute (dashed lines) and partial (solid lines)
correlations between the pressure $P$ (top) and energy density $\varepsilon$
(bottom) and the symmetry-energy parameters $J$, $L$, and $K_{\rm sym}$, all
evaluated at the same baryon density at which we want to quantify their relations.
The middle column displays the corresponding induced fractional variations
$\Delta P/P$ and $\Delta\varepsilon/\varepsilon$, with the shaded gray bands
indicating the total combined uncertainty obtained by adding in quadrature the
individual contributions from the isovector parameters. The right column presents the dominant symmetry-energy modes obtained from a
principal-component analysis of the local EOS response.
The principal-component
analysis (PCA) \cite{pca} is performed using the EOS sensitivities evaluated at all densities
within the inner crust, such that the resulting modes represent correlated
combinations of $J$, $L$, and $K_{\rm sym}$ governing the global crust EOS
response rather than density-local sensitivities.

While the absolute correlations of the pressure and energy density with the
isovector NM parameters are generally weak at different densities,
the influence of these parameters becomes significantly more apparent once
control variables are introduced through partial correlation analysis.
In contrast to the pressure, the energy density of the system is found to be
more strongly influenced by the isovector NM properties, particularly
at densities below $\rho \simeq 0.06~\mathrm{fm}^{-3}$. The partial correlation analysis reveals that fixing two of the three isovector
parameters exposes a strong conditional dependence of both the pressure and the
energy density on the remaining parameter, especially for the energy density.
In particular, the slope parameter $L$ shows a consistently stronger influence
on both $P$ and $\varepsilon$ in the low-density region of the inner crust,
although its effect remains comparable in magnitude to those of $J$ and
$K_{\rm sym}$. As can be seen from the correlations, no single NM
property dominates the EOS response at any density; instead, the sensitivities
arise from competing contributions of all three parameters.

In terms of magnitude, uncertainties in the isovector NM properties
can induce variations of up to $\sim\pm40\%$ in the pressure and
$\sim\pm20\%$ in the energy density at the lower inner crust densities.
These variations decrease substantially toward the crust--core transition
region, indicating a progressive suppression of isovector uncertainties at
higher densities. In the intermediate inner crust region
($\rho \lesssim 0.06~\mathrm{fm}^{-3}$), the slope parameter $L$ exhibits the
largest influence on both $P$ and $\varepsilon$, although not overwhelmingly so
compared to the other parameters, while $K_{\rm sym}$ generally shows the
smallest quantitative impact. To further quantify this collective behavior, we perform a PCA of the EOS response. The dominant symmetry-energy modes are found to be
\begin{align}
\delta P &\propto -0.466\,J - 0.668\,L + 0.581\,K_{\rm sym}, \\
\delta \varepsilon &\propto +0.613\,J + 0.534\,L - 0.583\,K_{\rm sym}.
\end{align}
These dominant modes emphasize that the inner crust EOS is governed by correlated
combinations of the symmetry-energy parameters rather than by any single nuclear
matter property. Notably, the effects of the symmetry-energy slope and curvature enter the EOS
with opposite signs for the pressure and energy density, leading to partial
cancellations in the total response when all isovector parameters vary
simultaneously. In particular, variations in $L$ and $K_{\rm sym}$ contribute
with opposite trends to $P$ and $\varepsilon$, which further suppresses the net
EOS uncertainty despite the strong conditional sensitivities revealed by the
partial correlation analysis.

\begin{figure*}
    \centering
    \includegraphics[width=1\linewidth]{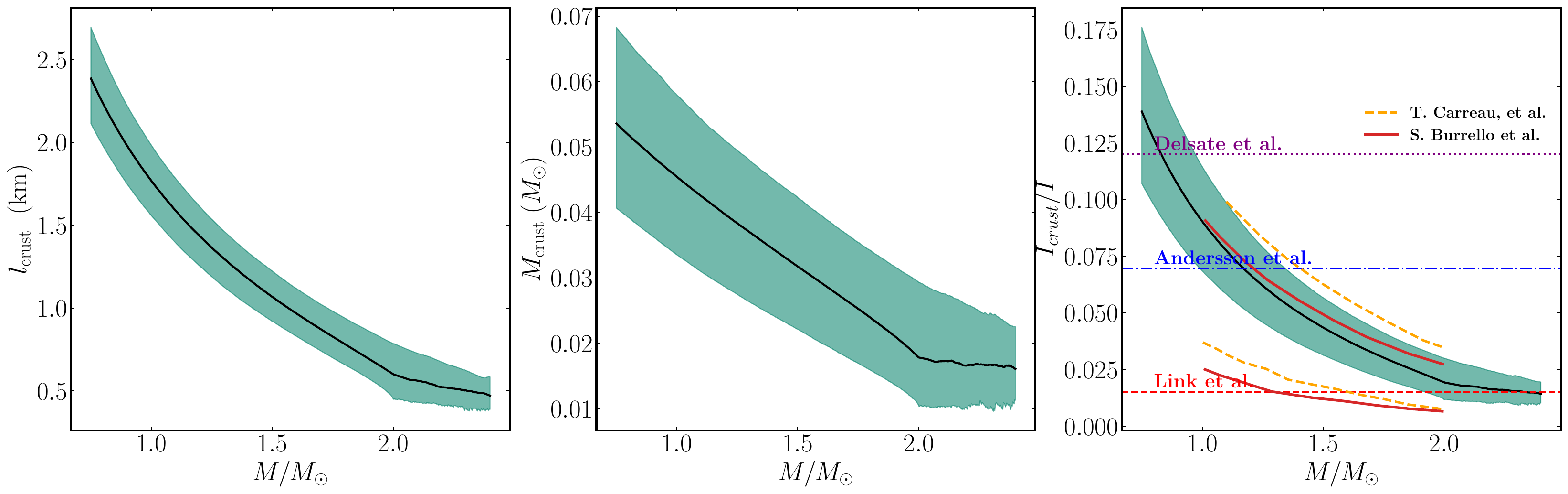}
    
    \caption{95\% credible interval bands for the crust thickness $L_{\rm crust}$,
the total crust mass $M_{\rm crust}$, and the crustal fraction of the moment of inertia
$I_{\rm crust}/I$. Results are shown for the case without superfluid entrainment
(solid band) and including entrainment effects using the prescriptions of
Link et al., Andersson et al., and Delsate et al. The comparison highlights the impact
of entrainment on the effective crustal angular-momentum reservoir.
}
    \label{fig:crust_tmi}
\end{figure*}

One of the most important observational probes of NS interiors is the occurrence of
pulsar glitches, which are  sudden spin-up events observed in several rotation-powered pulsars \cite{Wynn_2015, Pizzochero2017}. Among
these sources, the Vela pulsar provides the strongest constraint, exhibiting large and quasi-regular
glitches that require a substantial angular-momentum reservoir \cite{Pizzochero2017}. The standard interpretation
attributes these events to the sudden transfer of angular momentum from a neutron superfluid
in the inner crust to the solid crust and the rest of the star. For this mechanism to operate,
the crust must be sufficiently massive and thick so that it contains a large fraction of the total
moment of inertia. Detailed studies have shown that, depending on the crust--core transition
density and pressure, the crustal moment of inertia can reach a few percent and, in favorable
cases, approach the level required to explain Vela glitches even in the presence of superfluid
entrainment \cite{Carreau_2020, klausner2025neutronstarcrustinformed, Burrello_2025, Parmar_crust, Dinh_2021}. 

In Fig.~\ref{fig:crust_tmi}, we show the 95\% credible interval bands for the crust thickness
$l_{\rm crust}$, the total crust mass $M_{\rm crust}$, and the fractional crustal moment
of inertia as functions of the NS mass. All these properties are important parameter in various NS calculation such as glitches and  cooling \cite{Page_2004}.  Both the crust thickness and the crust
mass decrease monotonically with increasing stellar mass and show an approximately linear
trend over the considered mass range. Estimates of the crust thickness and mass obtained
from RMF models within the CLDM approach are reported
in Table~\ref{tab:crust_prop}.  The predicted crust thickness $l_{\rm crust}$ and total crust mass $M_{\rm crust}$ are comparable to the results of \cite{klausner2025neutronstarcrustinformed}, obtained using Skyrme energy density functionals within both the ETF and
CLDM approaches. Based on the present uncertainties in the
NS EOS arising from experimental nuclear observables and astrophysical observations, our results indicate that a NS with a mass of $1.4\,M_\odot$ is expected to have a crust thickness smaller than about $1.34$~km, with a corresponding crust mass below $0.045\,M_\odot$. This corresponds to  $10\%$ of the stellar radius and about $3.2\%$ of the total stellar mass. Furthermore, low-mass neutron stars ($M \lesssim 1.4,M_\odot$), whose global properties are already comparatively well constrained, provide a particularly powerful lever for probing the physics of the crust and, in turn, nuclear matter at sub-saturation densities. Owing to steady advances in experimental constraints on nuclear matter and parallel progress in theoretical modeling, the low- and intermediate-density regime of neutron-star matter is now tightly constrained compared to a decade ago. Looking ahead, high-precision mass–radius measurements of low-mass neutron stars, together with NICER-quality observations of canonical $\sim1.4,M_\odot$ pulsars, will offer complementary and highly sensitive information on the crustal mass and thickness, enabling significantly sharper tests of crust microphysics and its connection to nuclear matter properties.

\begin{table}
\centering
\caption{Median values and 95\% credible intervals for the crust thickness
$L_{\rm crust}$, total crust mass $M_{\rm crust}$, and fractional crustal
moment of inertia $ I_{crust}/I$ as functions of the NS mass.}
\label{tab:crust_props_ci}

\renewcommand{\arraystretch}{1.25}   
\setlength{\tabcolsep}{6pt}         

\begin{tabular}{cccc}
\hline
$M/M_\odot$ &
$l_{\rm crust}$ (km) &
$M_{\rm crust}$ ($M_\odot$) &
$ I _{crust} / I$ \\
\hline
$1.0$ &
$1.768^{+0.215}_{-0.210}$ &
$0.04544^{+0.01253}_{-0.01193}$ &
$0.0904^{+0.0242}_{-0.0219}$ \\

$1.2$ &
$1.434^{+0.180}_{-0.180}$ &
$0.03969^{+0.01125}_{-0.01109}$ &
$0.0667^{+0.0181}_{-0.0172}$ \\

$1.4$ &
$1.177^{+0.159}_{-0.158}$ &
$0.03435^{+0.01071}_{-0.01015}$ &
$0.0501^{+0.0145}_{-0.0137}$ \\

$1.6$ &
$0.967^{+0.151}_{-0.137}$ &
$0.02916^{+0.01033}_{-0.00917}$ &
$0.0378^{+0.0123}_{-0.0108}$ \\

$1.8$ &
$0.783^{+0.157}_{-0.125}$ &
$0.02393^{+0.01029}_{-0.00803}$ &
$0.0280^{+0.0106}_{-0.0088}$ \\

$2.0$ &
$0.601^{+0.185}_{-0.147}$ &
$0.01784^{+0.01150}_{-0.00738}$ &
$0.0193^{+0.0107}_{-0.0075}$ \\
\hline
\end{tabular}
\label{tab:crust_prop}
\end{table}

In Fig.~\ref{fig:crust_tmi}, we also show the posterior distributions of the fractional
crustal moment of inertia, $ I_{crust}/I$. We include the limiting case of zero entrainment
\cite{Link_1999} and scenarios with strong entrainment. The latter is commonly modeled by
rescaling the glitch activity parameter of the Vela pulsar by a prefactor of
$m_n^\ast/m_n \simeq 4.6$, following \cite{Andersson_2012}, as well as using the more
microscopic entrainment estimates of \cite{Delsate_2016}. The basic requirement for
explaining glitches within a crust-only superfluid framework is
\begin{equation}
\frac{I_{\rm crust}}{I} > \mathcal{G},
\label{eq:glitch_condition}
\end{equation}
where $\mathcal{G}$ is the dimensionless glitch activity parameter that quantifies the
fraction of the long-term spin-down reversed by glitches. For the Vela pulsar,
$\mathcal{G}_{\rm Vela} \simeq 0.016 \pm 0.002$ \cite{Wynn_2015}.

For the strongest entrainment scenario derived from the microscopic calculations of
\cite{Delsate_2016}, the condition in Eq.~\eqref{eq:glitch_condition} is not satisfied for
any NS mass, consistent with previous findings
\cite{klausner2025neutronstarcrustinformed,Carreau_2019}. This indicates that, if
entrainment is as strong as predicted by these models, a glitch mechanism relying solely
on crustal superfluidity is incompatible with observations. For the more widely adopted
strong-entrainment prescription with $m_n^\ast/m_n = 4.6$ \cite{Andersson_2012}, our RMF
posterior predicts that the condition $I_{\rm crust}/I > \mathcal{G}_{\rm Vela}$ can be
satisfied for NS masses below $\sim 1.3\,M_\odot$. This mass limit is slightly
higher than the value of $\sim 1.1\,M_\odot$ reported in
\cite{klausner2025neutronstarcrustinformed}. In contrast, in
the absence of entrainment, our RMF models satisfy Eq.~\eqref{eq:glitch_condition} over
nearly the entire mass range, failing only for very massive stars with
$M \gtrsim 2\,M_\odot$. Furthermore, when compared with the results of
\cite{Burrello_2025} and \cite{Carreau_2019}, we find that RMF-based crust models tend to
support the condition $I_{\rm crust}/I > \mathcal{G}_{\rm Vela}$ up to larger NS
masses. For completeness, Table~\ref{tab:crust_prop} also reports the 95\% credible interval for the
fractional crustal moment of inertia, $ I_{crust}/I$. For a canonical $1.4\,M_\odot$ neutron
star, we find a median value of $ I_{crust}/I \simeq 0.05$, indicating that the crust typically
stores about $5\%$ of the total stellar moment of inertia within the present uncertainties on NS matter. This value is comparable to the crustal fractional moment of inertia reported in Ref.~\cite{klausner2025neutronstarcrustinformed}, which was obtained within a Skyrme-based framework under various constraints on dense matter, suggesting a degree of robustness of this result across different nuclear interactions.

\begin{figure}
    \centering
    \includegraphics[width=1\linewidth]{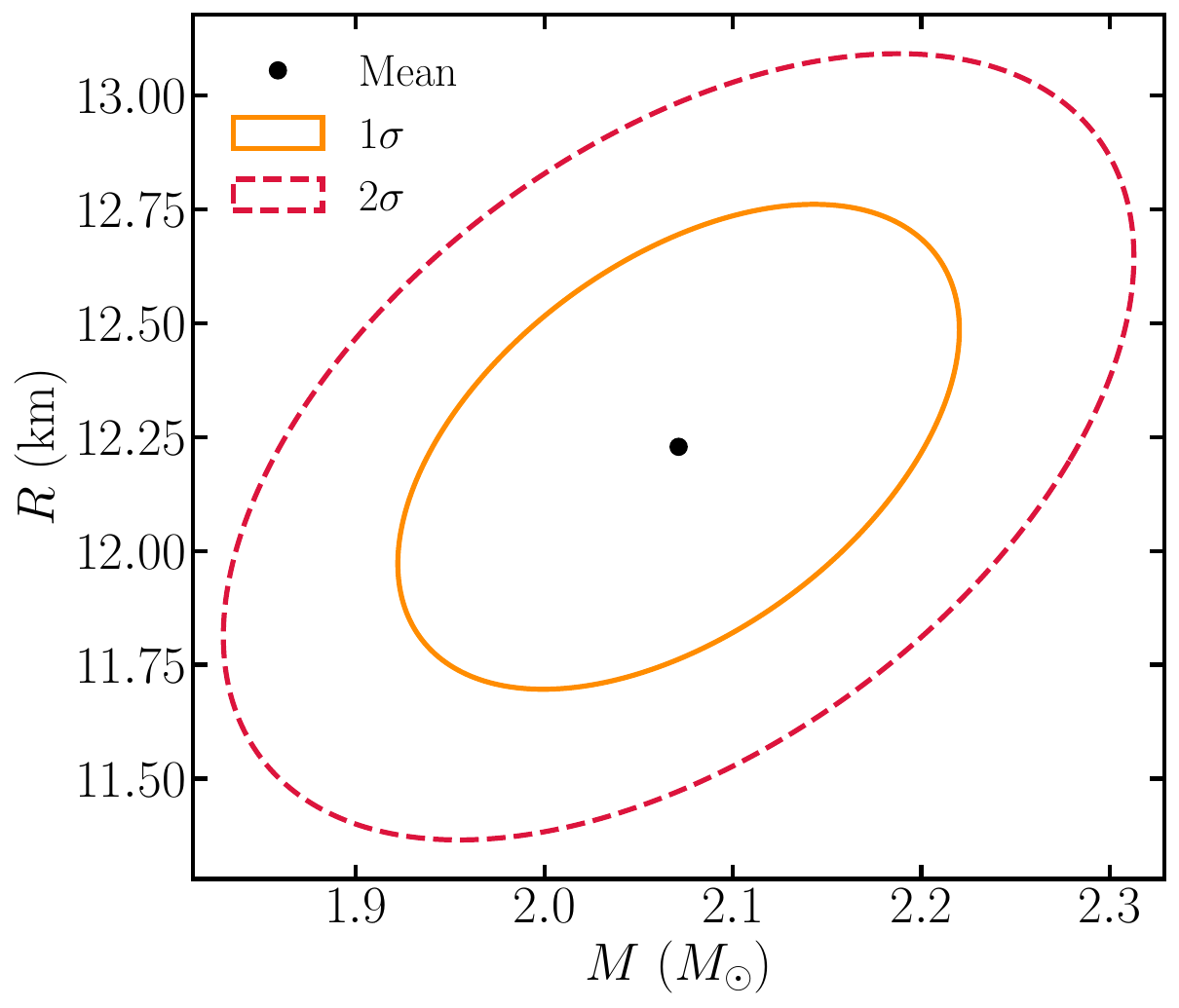}
\caption{Mass--radius constraints for the Vela pulsar . The solid and dashed ellipses denote the $1\sigma$ and $2\sigma$ confidence regions, respectively, constructed from the joint mass--radius probability distribution. The black marker indicates the mean inferred mass and radius.}
    \label{fig:mfromg}
\end{figure}

Regarding the Vela pulsar, its mass is not yet directly measured and is therefore uncertain. Estimates based on glitch activity and modeling suggest that its mass may lie in the range $1.5$--$1.8\,M_\odot$, as inferred in Ref.~\cite{Wynn_2015}. If such a high mass is assumed, then once strong neutron entrainment is taken into account, the crust alone is insufficient to account for the observed glitch activity.  However, in the absence of entrainment, the mass and radius of Vela can still be inferred from the measured glitch activity using our EOS ensemble. 
The M-R probability distribution can be  constructed as \cite{Carreau_2019_prc} 
\begin{equation}
p(M,R) = \sum_{\{\vec{P}_\alpha\}} p_{\rm post}(\{\vec{P}_\alpha\})
\,\delta(M_\alpha - M)\,\delta(R_\alpha - R),
\end{equation}
where \(M_\alpha\) and \(R_\alpha\) denote the mass and radius obtained for a given EOS parameter set \(\{\vec{P}_\alpha\}\). For each EOS realization, the stellar configuration is selected by requiring that the central density satisfies the glitch constraint. Here \(G\)  is sampled from a Gaussian distribution whose mean and variance are taken from \cite{Wynn_2015}. The resulting $1\sigma$ and $2\sigma$ confidence ellipses in the mass--radius plane are shown in Fig.~\ref{fig:mfromg}. Compared to Ref.~\cite{Wynn_2015}, our allowed region is significantly broader, since that study varied only the entrainment strength while keeping the underlying EOS fixed, whereas we marginalize over the full EOS uncertainty. Notably, the lower-mass edge of our confidence region overlaps with the largest Vela mass reported in Ref.~\cite{Wynn_2015}, $M = 1.83^{+0.04}_{-0.02}\,M_\odot$, indicating that uncertainties associated with neutron entrainment dominate over EOS uncertainties in setting the Vela glitch constraint. A similar conclusion was also reached in Ref.~\cite{Carreau_2019_prc}. Furthermore, in the absence of neutron entrainment, the inferred mass of the Vela pulsar is pushed toward relatively high values, exceeding $1.85\,M_\odot$ when the present uncertainties in neutron-star matter are taken into account.

\begin{figure*}
    \centering
    \includegraphics[width=.8\linewidth]{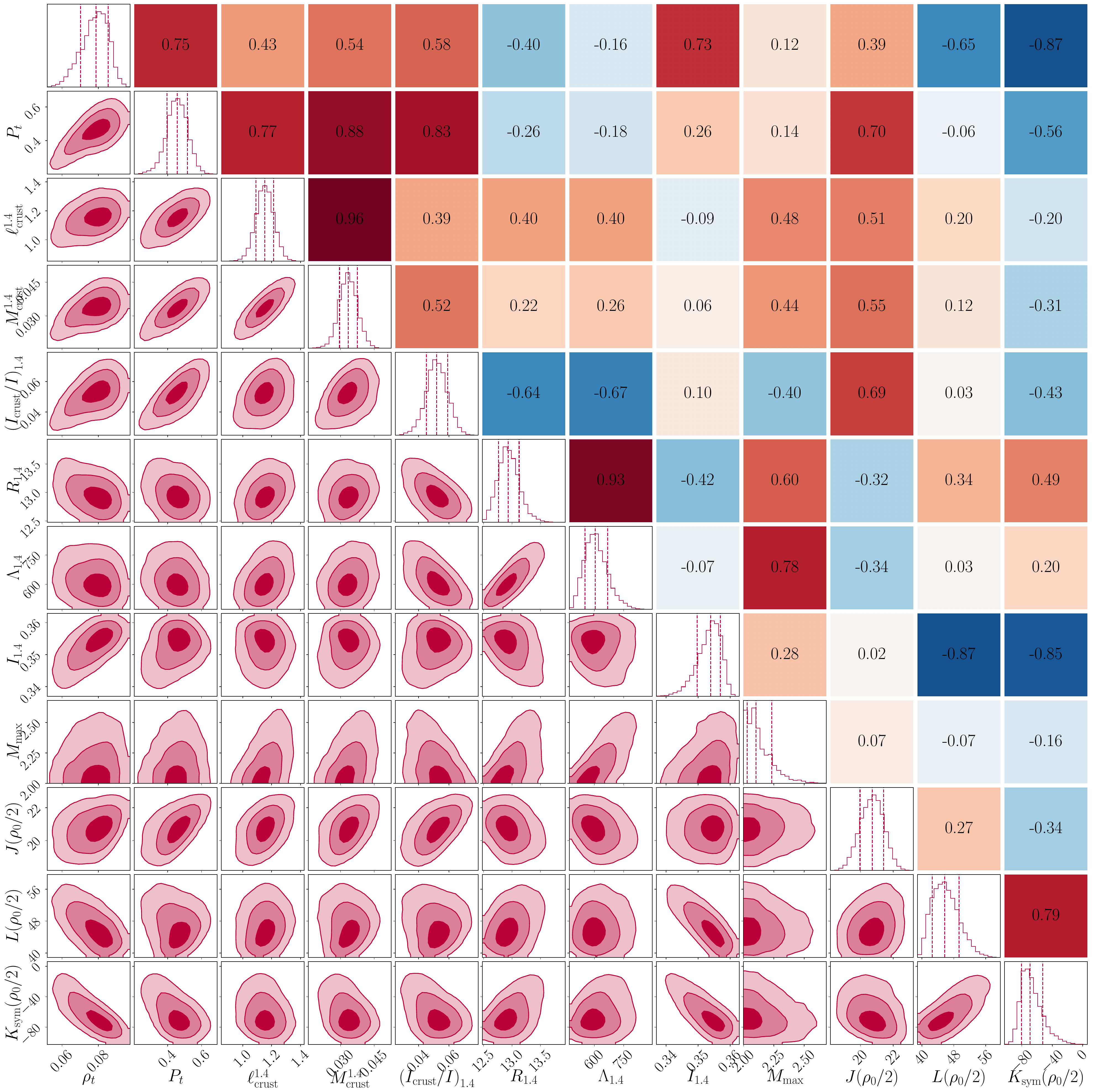}
    \caption{
Corner plot showing the joint posterior distributions of crustal properties
($\rho_t$, $P_t$, $l_{\rm crust} ^{1.4}$, $M_{\rm crust}^{1.4}$, $(I_{\rm crust}/I)_{1.4}$ ),
global NS observables ($R_{1.4}$, $\Lambda_{1.4}$, $I_{1.4}$, $M_{\rm max}$),
and NM parameters ($J$, $L$, $K_{\rm sym}$).
The diagonal panels display the marginalized one-dimensional posterior
distributions, while the lower-left panels show the two-dimensional
$68\%$ and $95\%$ credible regions.
The upper-right panels present the corresponding Pearson correlation
coefficients, color-coded to highlight positive (red) and negative (blue)
correlations.
}

    \label{fig:crust_property_all}
\end{figure*}

Finally, in Fig.~\ref{fig:crust_property_all}, we present the joint posterior
distributions of the crustal properties, namely,  the crust--core transition
density $\rho_t$, transition pressure $P_t$, crust mass $M_{\rm crust}$, crust
thickness $ l_{\rm crust}$, and fractional crustal moment of inertia
$I_{\rm crust}/I$, together with global NS observables
($R_{1.4}$, $\Lambda_{1.4}$, $I_{1.4}$, and $M_{\rm max}$) and NM
properties in the isovector channel ($J$, $L$, and $K_{\rm sym}$), evaluated at
$\rho_0/2$ rather than at saturation density. The figure reveals a rich pattern of correlations, among which several key
trends emerge. Most notably, the transition pressure $P_t$ is the dominant
crustal quantity, showing strong positive correlations with the crust
thickness, crust mass, and the fractional crustal moment of inertia. This
behavior reflects the central role of $P_t$ in setting the stiffness of the
inner crust and determining its macroscopic contribution to the stellar
structure. 

For a fixed stellar mass of $1.4\,M_\odot$, a thicker crust naturally leads to
a more massive crust, as evidenced by the tight correlation between
$l_{\rm crust}$ and $M_{\rm crust}$. At the same time, both the crust
thickness and mass display only weak or negligible direct correlations with
individual NM parameters, reinforcing the conclusion that crust
properties are governed by collective effects rather than by any single
parameter. The fractional crustal moment of inertia $I_{\rm crust}/I$ shows a clear
anticorrelation with the stellar radius $R_{1.4}$ and tidal deformability
$\Lambda_{1.4}$, illustrating the sensitivity of crustal observables to the
global stellar structure. This behavior suggests that future high-precision
radius and tidal measurements may provide indirect but meaningful constraints
on crust physics. Finally, evaluating NM properties at $\rho_0/2$ reveals additional
insights that are absent when they are defined at saturation density. In
particular, the fractional moment of inertia of a $1.4\,M_\odot$ NS
exhibits a noticeable correlation with $J$ when the latter is evaluated at
$\rho_0/2$, whereas this correlation is largely suppressed at $\rho_0$.
Moreover, the total moment of inertia $I_{1.4}$ is found to correlate with both
$L$ and $K_{\rm sym}$ at $\rho_0/2$, underscoring the importance of constraining
the EOS at sub-saturation densities for an accurate description
of NS structure.

\subsection{Unified EOS vs Matched EOS}
\begin{figure}
    \centering
    \includegraphics[width=1\linewidth]{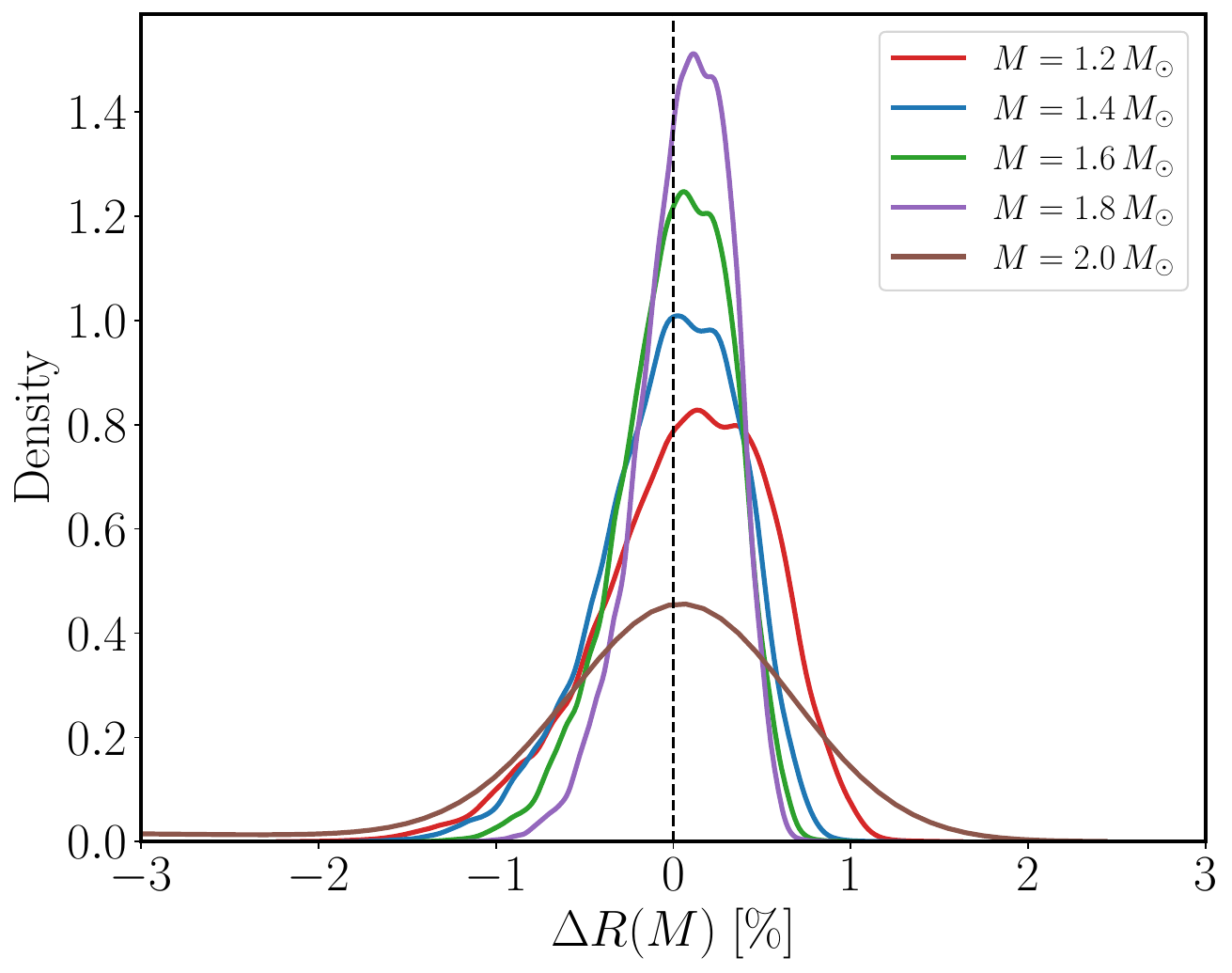}
    \caption{Kernel density estimates of the relative radius difference
$\Delta R(M) = [R_{\rm matched}(M)-R_{\rm unified}(M)]/R_{\rm unified}(M)$
for NSs with fixed masses
$M = 1.2, 1.4, 1.6, 1.8,$ and $2.0\,M_\odot$.
The vertical dashed line marks $\Delta R(M)=0$.
The distributions are obtained by comparing the radii from the unified
EOS construction with those from the original TOV solutions across the
full posterior ensemble.}
    \label{fig:matched_eos_radius}
\end{figure}
Finally, we discuss the implications of using unified EOS for
deducing the global structure of NSs. Although the crust accounts for
only a small fraction of the stellar mass, it plays a
non-negligible role in determining global observables such as the radius. As a
result, NS modeling often relies on non-unified EOS constructions, in
which the core EOS is matched to a separate crust EOS, most commonly based on
the BPS model for the outer crust and the Negele--Vautherin EOS for the inner
crust \cite{BPS,NEGELE1973298}. In such approaches, the crust--core transition is
implemented by enforcing continuity of pressure and chemical potential at a
chosen density, while the underlying nuclear interactions in the two regions
remain different. This widely used procedure introduces an intrinsic
uncertainty in the predicted stellar radius that depends on both the adopted
crust EOS and the matching prescription.

Studies have shown that different matching choices can lead to systematic
radius shifts at the level of a few percent, typically $\sim 2$--$5\%$, even
when the core EOS is fixed \cite{Fortin_2016, Suleiman_2021, Davis_2024}. 
With the steadily increasing precision of NS radius measurements from
NICER, which already constrain radii at the level of $\sim 0.5$--$1$~km for
sources such as PSR~J0030+0451 and PSR~J0740+6620, systematic uncertainties of
this magnitude are no longer negligible. This highlights the importance of
quantifying and controlling deviations introduced by matched EOS constructions,
and motivates the use of unified EOSs for robust interpretation of high-precision
NS observations. To quantify the uncertainty associated with unified and matched equations of
state, we compare the unified EOS obtained from our posterior with a matched
EOS constructed using a standard prescription. For the NS crust, we
carefully treat the interface regions. We employ the BPS EOS \cite{BPS} for the
outer crust, corresponding to energy densities in the range
$\varepsilon_{\rm min} \leq \varepsilon \leq \varepsilon_{\rm outer}$, where
$\varepsilon_{\rm min} = 1.0317\times10^{4}\,{\rm g\,cm^{-3}}$ and
$\varepsilon_{\rm outer} = 4.3\times10^{11}\,{\rm g\,cm^{-3}}$. In the density
region between the outer crust and the core,
$\varepsilon_{\rm outer} < \varepsilon \leq \varepsilon_c$, where
$\varepsilon_c$ denotes the crust--core transition density, we adopt a
polytropic EOS. The complete matched EOS is then defined as \cite{Chun_2024}
\begin{equation}
P(\varepsilon) =
\begin{cases}
P_{\rm BPS}(\varepsilon), & \varepsilon_{\rm min} \leq \varepsilon \leq \varepsilon_{\rm outer}, \\
A + B\,\varepsilon^{4/3}, & \varepsilon_{\rm outer} < \varepsilon \leq \varepsilon_c, \\
P_{\rm RMF}(\varepsilon), & \varepsilon > \varepsilon_c ,
\end{cases}
\end{equation}
where $P_{\rm RMF}$ denotes the pressure obtained from the RMF framework
introduced previously. The parameters $A$ and $B$ are chosen to ensure
continuity of the pressure at both $\varepsilon_{\rm outer}$ and
$\varepsilon_c$. The resulting inner crust EOS smoothly interpolates
between the outer crust and the core EOS and is finally matched to the core at
$\varepsilon_c \simeq 2.14\times10^{14}\,{\rm g\,cm^{-3}}$.

In Fig.~\ref{fig:matched_eos_radius}, we show the relative difference in the stellar
radius between matched and unified EOS for NS masses
ranging from $1.2\,M_\odot$ to $2.0\,M_\odot$. We find that lower-mass NSs
exhibit the largest spread in radius deviations, indicating that a larger fraction
of EOSs display noticeable differences when a matched construction is adopted
instead of a unified one. In this mass range, the radius deviation can reach the
level of $\sim 2\%$. As the stellar mass increases, the spread in the radius difference progressively
narrows, reflecting the diminishing influence of the crust on the global stellar
structure. However, for EOSs supporting masses close to $2\,M_\odot$, the band
broadens again. This behavior is a direct consequence of the matching procedure
employed here, in which the crust--core transition is imposed at a fixed density
rather than being determined self-consistently from the underlying nuclear
interaction. Different matching prescriptions, as discussed above, may therefore
lead to quantitatively different levels of deviation when compared to a fully
unified EOS. Importantly, crust-induced deviations are not confined to low-mass NSs
but propagate systematically to higher-mass configurations, underscoring the need
for an accurate treatment of the crust even in heavy NSs. This becomes
particularly relevant in light of the increasing precision of radius measurements
from \textit{NICER} and future multi-messenger analyses, where percent-level
systematic effects can no longer be neglected. Although the absolute difference in radius remains at the level of a few percent,
its impact on the tidal deformability is significantly amplified due to the strong
scaling $\Lambda \propto R^5$. Consequently, even modest crust-induced radius
variations can lead to appreciable differences in inferred tidal deformabilities,
as discussed in Ref.~\cite{Fortin_2016}. Since both the crust mass and thickness are
highly sensitive to the crust--core transition density, these results further
emphasize the importance of a unified treatment of the EOS for
robust and consistent NS modeling.

\section{\label{conclusion} Summary and Outlook}

In this work, we perform a Bayesian analysis of neutron-star crust properties using a unified relativistic mean-field (RMF) description of dense matter, incorporating constraints from nuclear experiments, chiral effective field theory, and multimessenger neutron-star observations. Unlike
most previous studies, which primarily constrain the uniform core and attach an
external crust model, our approach treats the outer crust, inner crust, and
liquid core consistently within the same microscopic interaction.

For each  sample, the outer crust was constructed using the AME2020
nuclear mass evaluation supplemented by HFB mass tables, while the inner crust
was modeled using a compressible liquid-drop model whose surface and curvature
parameters were refitted consistently to nuclear masses. The resulting ensemble
of unified EOS spans the full density range relevant for neutron
stars and satisfies all imposed nuclear and astrophysical constraints.

We obtained  predictions for key crust properties, including
the crust--core transition density and pressure, crust thickness, crust mass,
and the fractional crustal moment of inertia. Our results confirm that the
crust--core transition is primarily governed by isovector NM
properties, in particular the symmetry energy slope and curvature, rather than
by the symmetry energy at saturation or the incompressibility. Evaluating nuclear
matter properties at sub-saturation densities, around $\rho_0/2$, significantly
strengthens the observed correlations and provides a more physically relevant
description of crust physics.

A detailed sensitivity analysis of the inner-crust EOS showed that
its behavior is intrinsically collective. Absolute correlations between pressure
or energy density and individual NM parameters are generally weak,
but partial correlation analysis reveals strong conditional dependencies once
correlations among symmetry-energy parameters are accounted for. No single
NM parameter dominates the crust EOS at any density;
instead, the pressure and energy density arise from competing and partially
canceling contributions from the symmetry energy, its slope, and its curvature.

We further quantified the impact of crust physics on global NS
observables. The transition pressure was found to be the dominant crustal
quantity, strongly correlated with crust thickness, crust mass, and the
fractional crustal moment of inertia. For a canonical $1.4\,M_\odot$ neutron star, our unified models predict that the crust typically contains of order $5\%$ of the total stellar moment of inertia. The corresponding crustal thickness is found to be $\ell_{\rm crust} = 1.177^{+0.159}_{-0.158}\,\mathrm{km}$, while the crustal mass amounts to $M_{\rm crust} = 0.03434^{+0.0107}_{-0.0101}\,M_\odot$.
 When superfluid entrainment is
included, a crust-only explanation of large pulsar glitches remains viable only
for relatively low-mass NSs, consistent with earlier studies.

Finally, we explicitly compared unified and matched EOS and
demonstrated that commonly used matching prescriptions introduce systematic
radius differences at the level of a few percent. While modest in absolute
terms, these differences can be amplified in tidal deformabilities and therefore
become non-negligible in the context of current high-precision multimessenger
constraints. Our results highlight the importance of unified EOS
for reducing systematic uncertainties and for a consistent interpretation of
NS observations.

In addition to global crustal properties, microscopic features of the inner crust, such as the shear modulus, its spatial variation, and the possible emergence of nuclear pasta phases, play a central role in shaping the spectrum of crustal oscillations. These ingredients directly affect the frequencies of torsional shear modes and are therefore crucial for interpreting the quasi-periodic oscillations observed in magnetar flares, which provide one of the most direct observational probes of the neutron-star crust and its internal structure. A consistent treatment of the shear response and pasta phases within a unified equation-of-state framework is thus essential for linking theory to observations. We will pursue a detailed investigation of these effects in future work.

\bibliography{crust}

\end{document}